\documentclass[aip,apl,reprint,final]{revtex4-1}

\usepackage{graphicx,amsmath,amssymb,mathrsfs,array,longtable,delarray}
\usepackage{dcolumn,bm} 

\begin{document}

\title{Ultra-shallow quantum dots in an undoped GaAs/AlGaAs 2D electron gas}

\author{W. Y. Mak}
\author{F. Sfigakis}
\affiliation{Cavendish Laboratory, University of Cambridge, Cambridge, United
Kingdom}
\author{K. Das Gupta}
\affiliation{Cavendish Laboratory, University of Cambridge, Cambridge,
United Kingdom} \affiliation{Department of Physics, Indian Institute of
Technology Bombay, Mumbai 400076, India}
\author{O. Klochan}
\affiliation{School of Physics, University of New South Wales, Sydney, Australia}
\author{H. E. Beere}
\affiliation{Cavendish Laboratory, University of Cambridge, Cambridge,
United Kingdom}
\author{I. Farrer}
\affiliation{Cavendish Laboratory, University of Cambridge, Cambridge,
United Kingdom}
\author{J. P. Griffiths}
\affiliation{Cavendish Laboratory, University of Cambridge, Cambridge,
United Kingdom}
\author{\\G. A. C. Jones}
\affiliation{Cavendish Laboratory, University of Cambridge, Cambridge,
United Kingdom}
\author{A. R. Hamilton}
\affiliation{School of Physics, University of New South Wales, Sydney, Australia}
\author{D. A. Ritchie}
\affiliation{Cavendish Laboratory, University of Cambridge, Cambridge, United
Kingdom}

\begin{abstract}
We report quantum dots fabricated on very shallow 2-dimensional electron
gases, only 30\,nm below the surface, in undoped GaAs/AlGaAs
heterostuctures grown by molecular beam epitaxy. Due to the absence of
dopants, an improvement of more than one order of magnitude in mobility
(at 2$\times$10$^{11}$\,cm$^{-2}$) with respect to doped heterostructures
with similar depths is observed. These undoped wafers can easily be gated
with surface metallic gates patterned by e-beam lithography, as
demonstrated here from single-level transport through a quantum dot
showing large charging energies (up to 1.75\,meV) and excited state
energies (up to 0.5\,meV).
\end{abstract}


\maketitle

Electrostatically-defined quantum dots fabricated on high-mobility
GaAs/AlGaAs heterostructure have been $-$ and continue to be $-$
invaluable in many fundamental investigations, \textit{e.g.} Kondo physics
and spin-based solid-state qubits. Unfortunately, the characteristics of
these devices are extremely sensitive to seemingly random charge
fluctuations in their local electrostatic potential, commonly known as
Random Telegraph Signal (RTS) noise, or charge noise. Although one can
perform a biased cooling\cite{Pioro05-B,Buizert08} or a thermal
cure\cite{Fujiwara02} to attempt to drastically reduce the levels of
charge noise on a given device, results from both techniques vary from
device to device.

Quantum dots fabricated in shallow two-dimensional electron gases (2DEGs) have two
advantages over their deeper cousins. First, finer features can be transferred from
the surface metallic gates to the 2DEG. Second, the energy scales of the dot levels
tends to be larger, which enable operation at higher temperatures. However, shallow
2DEG depths (as little as 20\,nm below the surface) come at the expense of
mobility.\cite{Kopnov10,Laroche10,Giesbers10,Rossler10-B,Nemutudi05,Gildmeister07,
Nakamura10,Muhle08,Fricke07,Nemutudi04,Goldhaber98-A} Furthermore, the dopant layer
may partially screen surface gates (through hopping conduction) and/or facilitate
gate leakage, rendering many such wafers ungateable by surface metallic gates. The
ungateability of some doped wafers is not only restricted to shallow 2DEGs, but also
can occur in high-mobility doped wafers.\cite{Miller07,Dolev08,Rossler10-A}

The limitations described above can be circumvented or mitigated by using
undoped heterostructures in different field-effect transistor (FET)
geometries such as the
SISFET\cite{Kane93,Kawaharazuka01,Reilly02,Noh03-A,Lilly03}
(semiconductor-insulator-semiconductor), the
MISFET\cite{Herfort96,Harrell99,Willett06}
(metal-insulator-semiconductor), or the HIGFET\footnote{The term HIGFET
has been used in the literature to refer to both SISFET and/or MISFET
geometries. By MISFET, we mean the gate is a metal (e.g. Au) and the
insulator is some dielectric. By SISFET, we mean the gate is a
highly-doped semiconductor (metallic regime) and the insulator is also a
semiconductor.} (heterostructure-insulator-gate). Since there are no
intentional dopants, the 2DEG can be brought much closer to the surface
without sacrificing mobility. Furthermore, undoped quantum dots would not
suffer from one possible source of charge noise: electrons hopping between
dopant sites in AlGaAs. In doped wafers, intentional dopants typically
outnumber unintentional dopants 10,000 to 1 (depending on mobility).
Finally, undoped quantum dots may also interact with fewer undesirable
impurities in the vicinity and are far more reproducible upon thermal
cycling than their doped counterparts.\cite{See12} In this Letter, we
compare ultra-shallow undoped and doped GaAs/AlGaAs 2DEGs, and demonstrate
gated quantum dots on ultra-shallow undoped heterostructures.

\begin{figure}[t]
    \begin{minipage}{0.85\columnwidth}
    \includegraphics[width=\columnwidth]{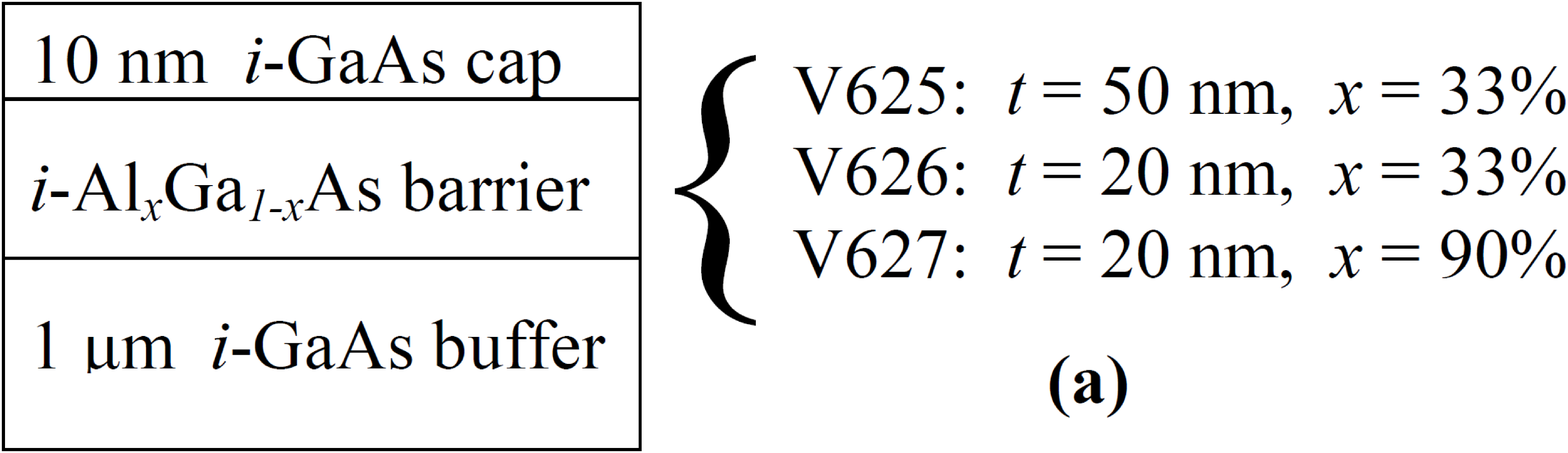}
    \vspace{0.05cm}
    \end{minipage}
    \begin{minipage}{0.46\columnwidth}
    \includegraphics[width=\columnwidth]{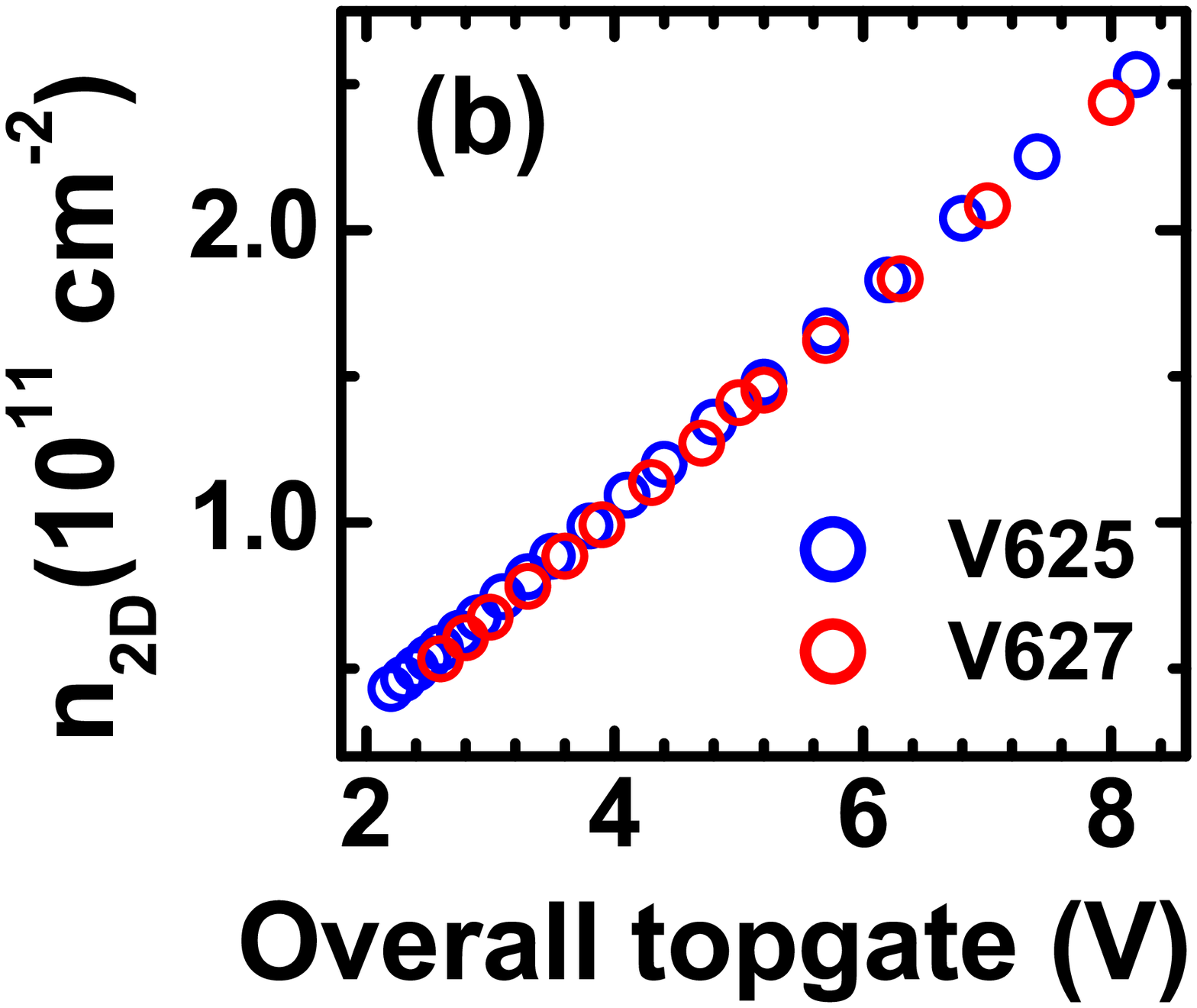}
    \end{minipage}
    \hspace{0.01cm}
    \begin{minipage}{0.51\columnwidth}
    \includegraphics[width=\columnwidth]{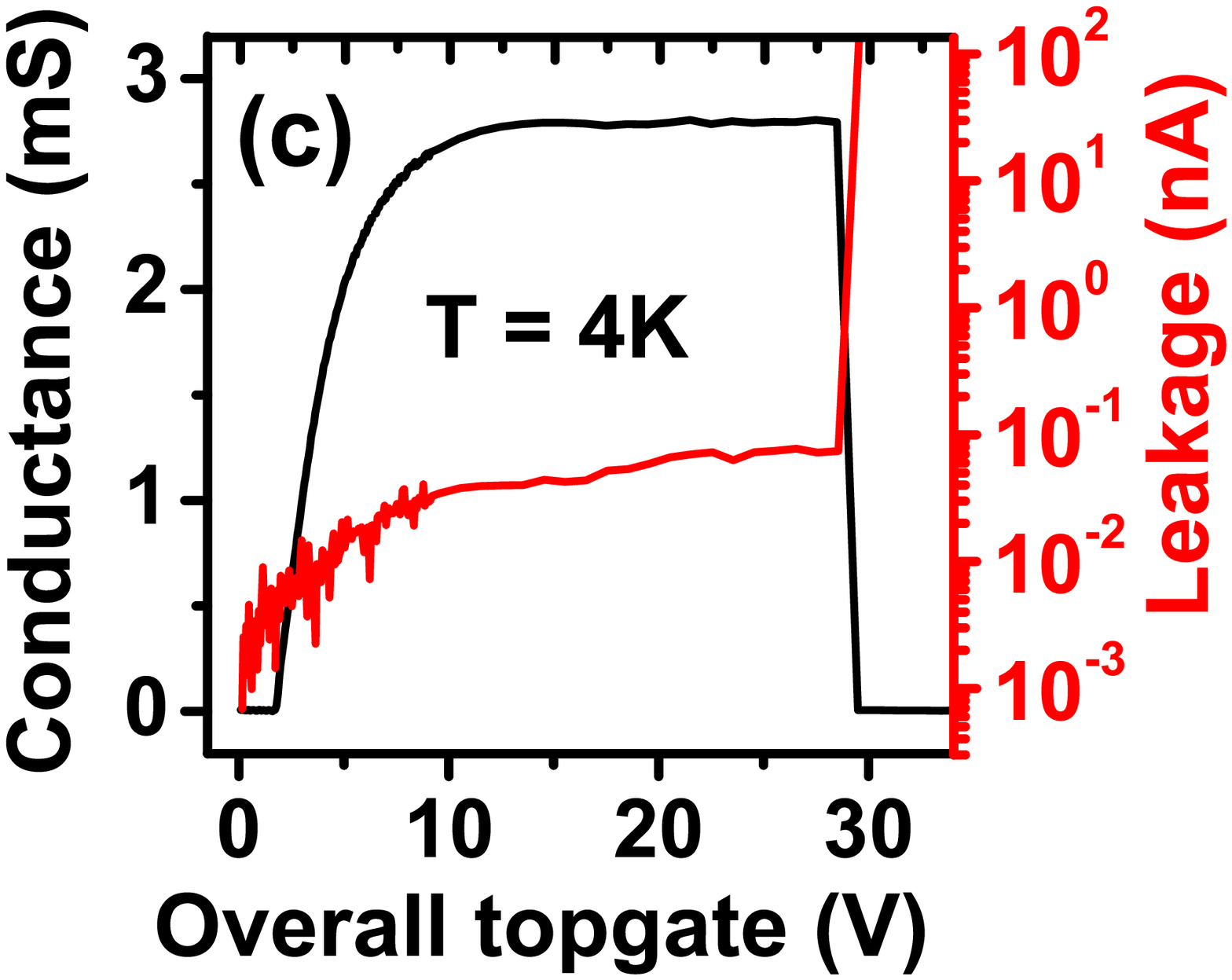}
    \end{minipage}
    \begin{minipage}{1.00\columnwidth}
    \vspace{0.2cm}
    \includegraphics[width=\columnwidth]{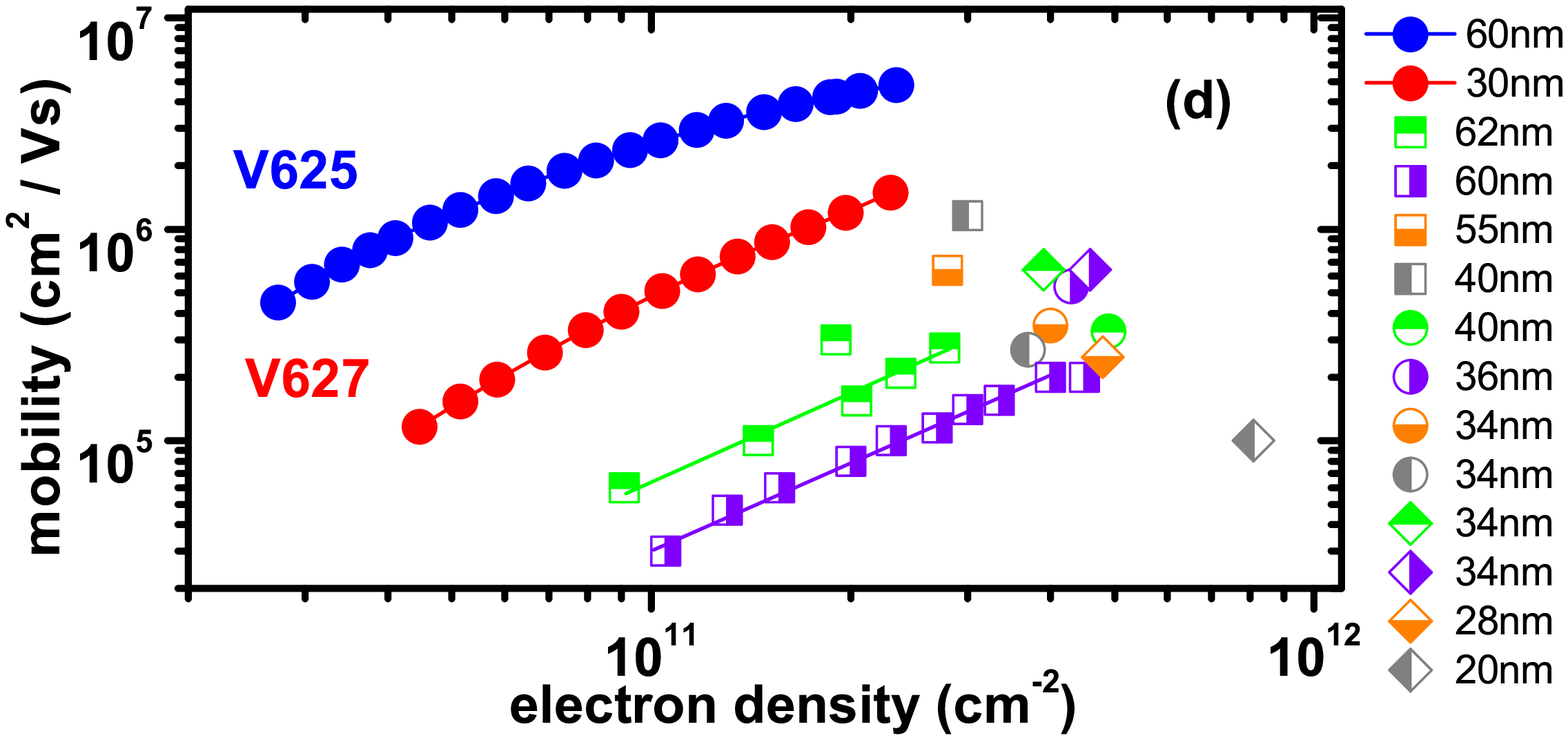}
    \end{minipage}
    \caption{(a) MBE layers used in the undoped heterostructures.
    (b) Carrier density versus overall topgate voltage for the V625 and V627
    undoped 2DEGs. The capacitance is dominated by the 500nm-thick polyimide layer.
    (c) Leakage current (right) to the 2DEG as the overall topgate is swept, with the
    accompanying two-terminal conductance between a pair of ohmic contacts (left).
    (d) Mobility versus electron density for undoped wafers V625 and V627,
    as well as various other doped wafers in shallow 2DEGs.\cite{Kopnov10,Laroche10,
    Giesbers10,Rossler10-B,Nemutudi05,Gildmeister07,Nakamura10,Muhle08,Fricke07,
    Nemutudi04,Goldhaber98-A} Lines through data points indicate a gated measurement.
    The legend indicates the 2DEG depth below the surface.}
    \label{fig1}
\end{figure}

Three undoped heterostructures were grown on the same day (V625, V626, and
V627) with their MBE layers shown in Figure \ref{fig1}(a). Details of
fabrication are otherwise identical to and extensively described in
Ref.\,\onlinecite{Wendy10}. The surface Ti/Au gates defining our quantum
dots were fabricated by e-beam lithography directly on the surface of the
GaAs cap, below the insulator layer (500\,nm of polyimide, or 175\,nm of
SiO$_2$). Above the insulator layer, a Ti/Au overall topgate covers the
entire surface of the 2DEG (overlapping the ohmic contacts) and varies the
carrier density. Measurements were performed in a pumped-{$^3$}He
cryostat.

The typical failure mechanism in SISFETs and MISFETs are electrical shorts
between the ohmic contacts and the overall topgate.\cite{Sarkozy07} None
of our samples suffered from this problem. Indeed, the relationship
between density and the overall topgate voltage is linear for V625 and
V627 [Fig.\,\ref{fig1}(b)], and the carrier density does not saturate with
overall topgate voltage. Both observations are consistent with no leakage
from the overall topgate to the 2DEG [Fig.\,\ref{fig1}(c)]. However, we
found that in V626 the AlGaAs barrier was not insulating enough to prevent
charge leaking between the 2DEG and the semiconductor-insulator interface
(a GaAs-polyimide interface in our case). Shortly after the 2DEG is fully
induced, charge ``seeps'' through the AlGaAs barrier. As it gradually
accumulates at the GaAs-polyimide interface, it begins to screen the
overall topgate from the 2DEG, which results in a gradual loss of carriers
until the 2DEG pinches off with time.\footnote{The timescale for this
process is of order of a several minutes, with factors being the
resistance and capacitance across the AlGaAs barrier (thus determining an
{$RC$} time constant), and the speed at which the overall topgate is
ramped (or if held steady). In this scenario, we emphasize that no leakage
is observed between the 2DEG and the overall topgate: no current flows
through the insulating polyimide layer.} In V627, increasing the energy
height of the Al$_{x}$Ga$_{1-x}$As barrier (by using $x$\,=\,0.90 instead
of $x$\,=\,0.33) eliminated the ``seeping'' leak between the 2DEG and the
GaAs-polyimide interface. The ultra-shallow 2DEG in V627 was stable in
time and did not leak to the overall topgate [Fig.\,\ref{fig1}(c)], and
neither do any of our deeper undoped 2DEGs.

\begin{figure}[t]
    \begin{minipage}{0.46\columnwidth}
    \includegraphics[width=\columnwidth]{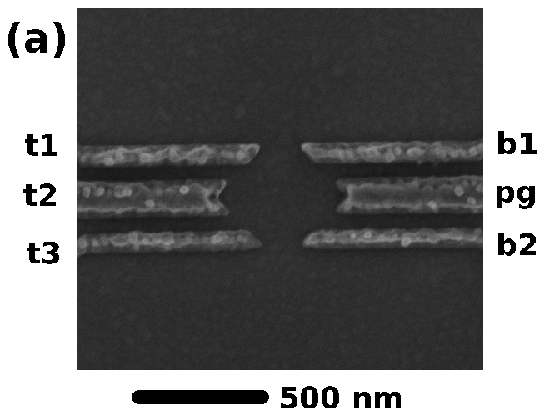}
    \end{minipage}
    \hspace{0.2cm}
    \begin{minipage}{0.45\columnwidth}
    \includegraphics[width=\columnwidth]{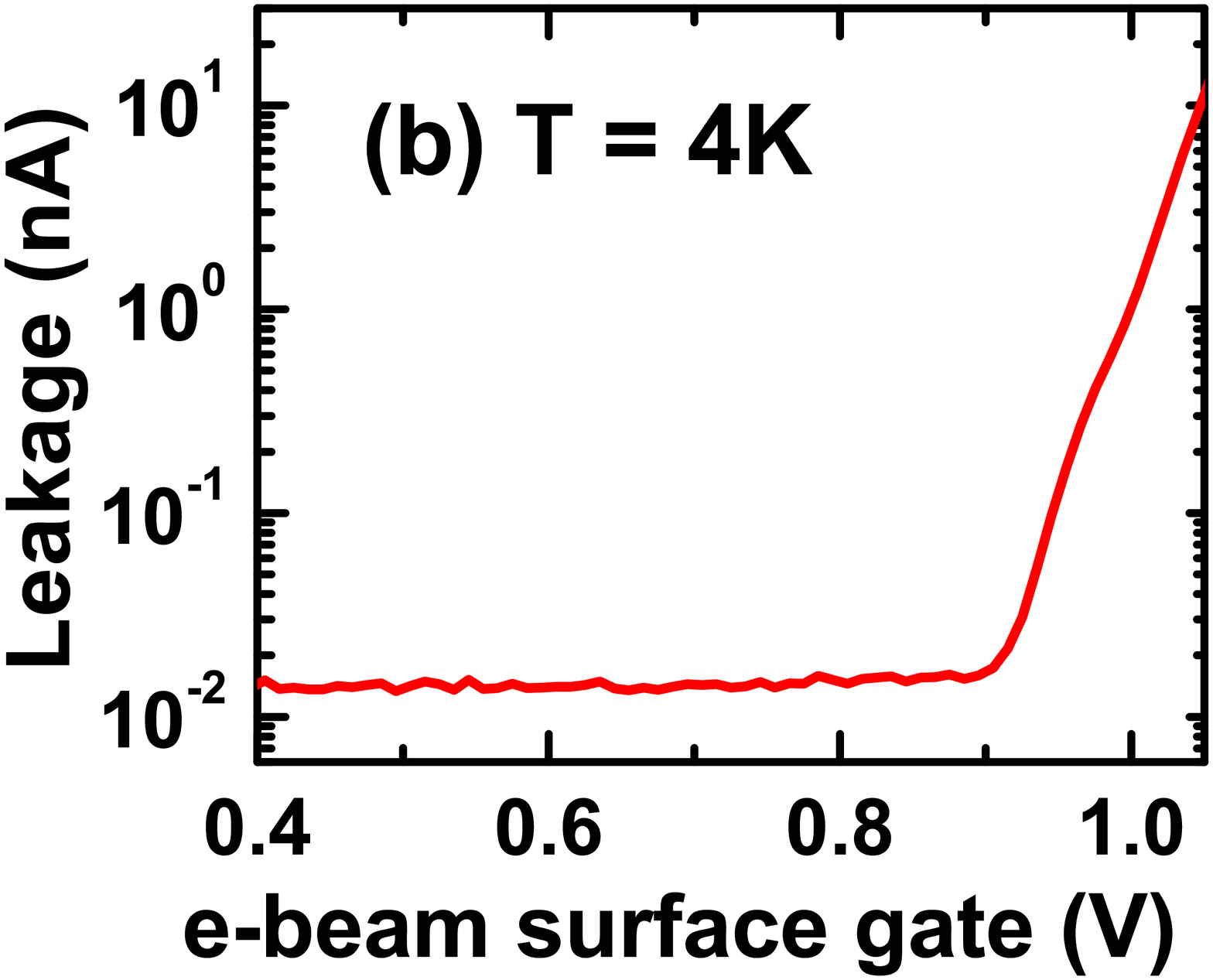}
    \end{minipage}
    \begin{minipage}{1.0\columnwidth}
    \includegraphics[width=\columnwidth]{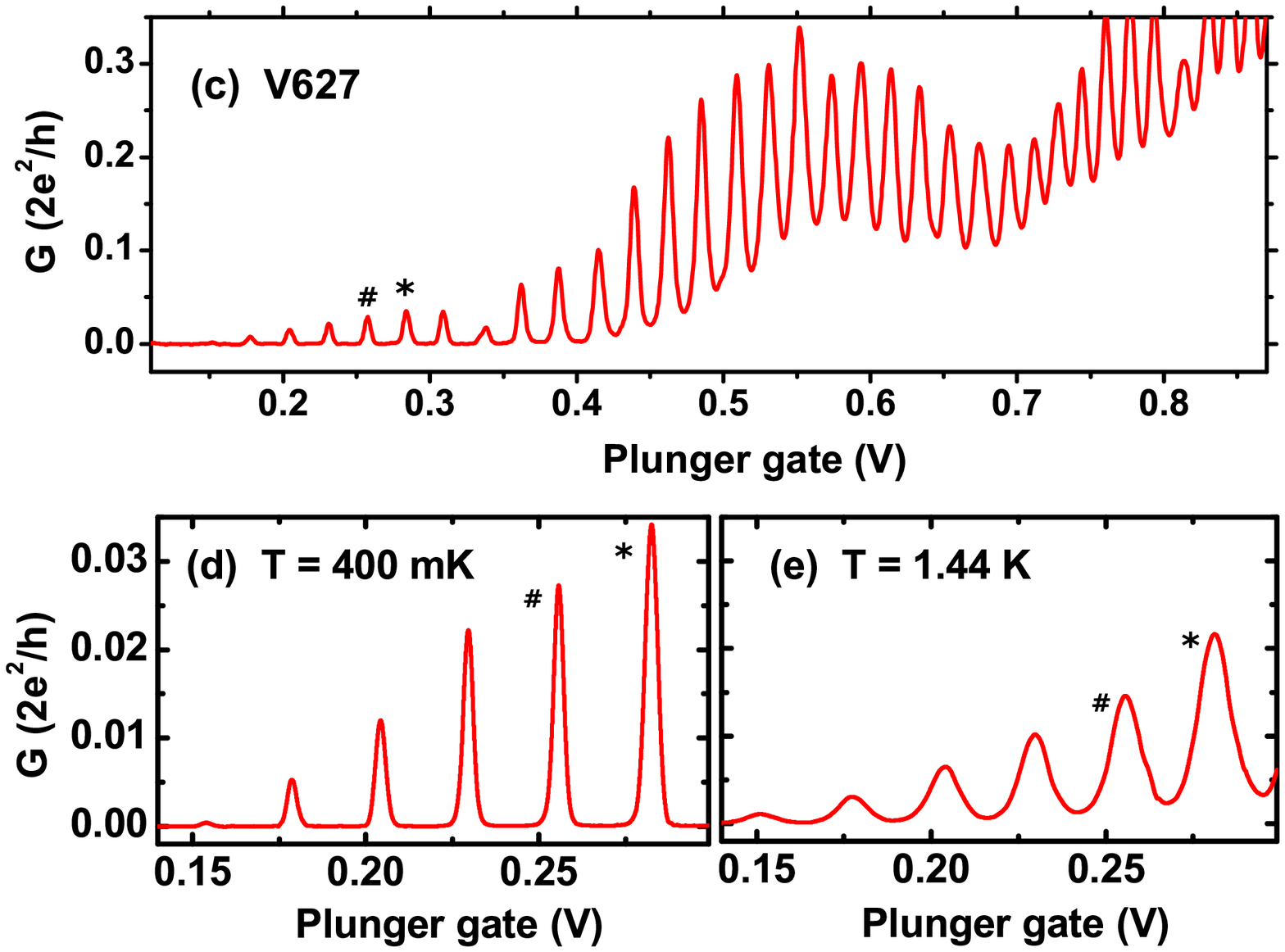}
    \vspace{0.01cm}
    \end{minipage}
    \caption{(a) Micrograph of a dot with the same dimensions as the one reported
    here. The surface gates t1, t2, and t3 are held together at the same voltage. Gates
    b1 and b2 control the left and right barriers, and gate pg is the plunger.
    (b) Leakage characteristics of surface gates defined by e-beam lithography,
    in another similar device.
    (c) Measured differential conductance $G$ (with 20\,$\mu$V AC excitation) showing
    Coulomb blockade (CB) oscillations from a quantum dot on wafer V627. All CB
    data was obtained at an overall topgate voltage of 8.0V
    ($n=2.5\times10^{11}$ cm$^{-2}$). Gates t1, t2, t3, b1, and b2 are all held at
    0.900V while pg is swept. CB peaks marked with $\#$ and $\ast$ are further
    analyzed in Figure \ref{fig4}.
    (d) The dot is in the weak coupling regime.
    (e) Even at high temperature, the CB peaks are well resolved, indicative of a
    large charging energy.}
    \label{fig2}
\end{figure}

Figure \ref{fig1}(d) compares 2D transport characteristics between shallow
GaAs/AlGaAs 2DEGs from our undoped wafers and from a representative sample of
published values of doped wafers. The large improvement in mobility (more than one
order of magnitude at 2$\times$10$^{11}$\,cm$^{-2}$) is largely attributable to the
absence of intentional dopants. Although scattering due to surface states is
significant for a 2DEG depth less than 80\,nm from the surface,\cite{Wendy10} it is
still by far the scattering from remote ionized impurities that impairs mobility in
doped shallow 2DEG wafers.\cite{Gold88,MacLeod09} The difference in mobility between
V625 and V627 is mostly accounted for by surface states which we simulated as a
delta-doped layer at the surface for the purposes of scattering.\cite{Wendy10}

It is only recently that quantum dots have been fabricated in undoped
heterostructures with electrons\cite{See10} and holes.\cite{Klochan10} No
RTS event could be observed in the electron quantum dot fabricated on the
SISFET of Ref.\,[\onlinecite{See10}]. Figure \ref{fig2} shows Coulomb
blockade (CB) oscillations in the weak coupling regime from one of three
quantums dot fabricated on wafer V627 (the other two devices have similar
characteristics as those shown here). The surfaces gates defining the
quantum dot [Fig.\,\ref{fig2}(a)] do not leak significantly below
$\sim$0.95V [Fig.\,\ref{fig2}(b)]. Our typical leakage current is
10-20\,pA, the effective sensitivity floor of our measurement setup. All
surface gate voltages are positive: they partially screen the overall
topgate and thus require a positive bias to help the overall topgate
induce a 2DEG at the center of the quantum dot. Directly underneath the
surface gates, the electric field from the overall topgate is totally
screened and a 2DEG does not form until a much higher positive voltage is
applied to the surface gates.

\begin{figure}[t]
    \includegraphics[width=\columnwidth]{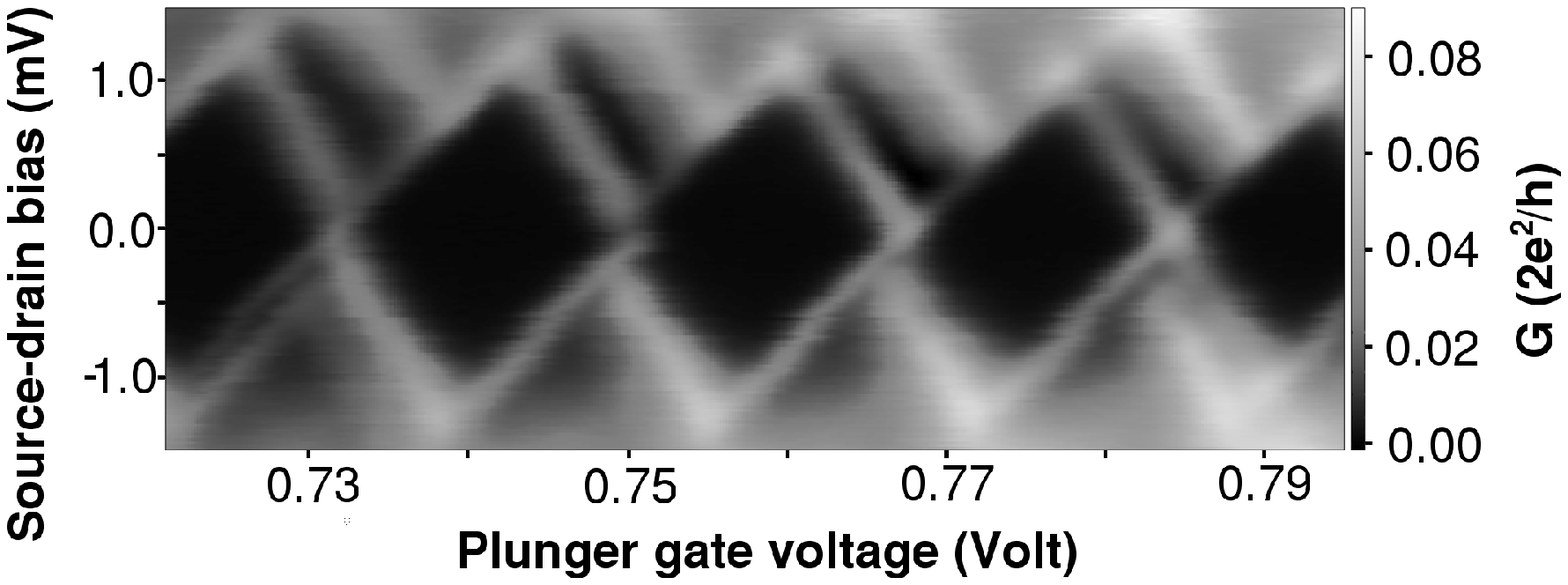}
    \caption{Source-drain bias spectroscopy. The dot is in a slightly different regime
    than shown in Figure \ref{fig2}: gates t1\,=\,t2\,=\,t3\,=\,0.836V, b1\,=\,0.862V,
    b2\,=\,0.918V, and the overall topgate at 8V. Parameters extracted from this data
    set are: the charging energy ($U$\,=\,1.25\,meV), the single-particle energy level
    spacing ($\Delta E$\,=\,0.5\,meV), the plunger gate lever arm
    ($e\alpha$\,=\,0.083\,meV/mV), the plunger capacitance ($C_g$\,=\,$e/\Delta V_g$
    =\,9.5\,aF), and the total capacitance ($C_\Sigma$\,=\,128\,aF). }
    \label{fig3}
\end{figure}

Despite its large dimensions, the charging energy ($U$\,=\,$e^2/C_\Sigma$,
where $C_\Sigma$ the total capacitance) of our our quantum dot is rather
large: $U$\,=\,1.25\,meV [Fig.\,\ref{fig3}]. By suitably changing gate
voltages, $U$ can be tuned as high as 1.75\,meV, but the dot is then no
longer in the weak coupling regime. The (weak coupling regime) charging
energy in our MISFET ($U$\,=\,1.25\,meV) is larger than that
($U$\,=\,0.5\,meV) of the quantum dot on the SISFET from
Ref.\,[\onlinecite{See10}]. The single-particle energy level spacing
($\Delta E$) of our quantum dot is also larger, $\Delta
E$\,$\sim$\,0.5\,meV as opposed to $\Delta E$\,$\sim$\,0.2\,meV, and
enables the observation of excited states at 400\,mK. The total
capacitance of our MISFET dot (128\,aF) is slightly less than that of the
SISFET dot (160\,aF), and is dominated by the capacitance to the large
overall topgate above the quantum dot in both cases. Using the 2D electron
density and the area implied by the measured total capacitance (assuming
$C_\Sigma$\,=\,$\varepsilon A/d$), there are at most $\sim\,$80 electrons
in our dot. The energy scales of $U$ and $\Delta E$ are directly related
to the 2DEG depths of the SISFET (185\,nm deep) and our MISFET (30\,nm
deep). Using $C_\Sigma$\,=\,$\varepsilon A/d$ and the values above, we
estimate a quantum dot fabricated on a hypothetical SISFET with a 30\,nm
deep 2DEG would have a large total capacitance, 800$-1000$\,aF. It
therefore appears more advantageous to fabricate quantum dots from MISFETs
rather than SISFETs.

\begin{figure}[t]
    \begin{minipage}{0.47\columnwidth}
    \includegraphics[width=\columnwidth]{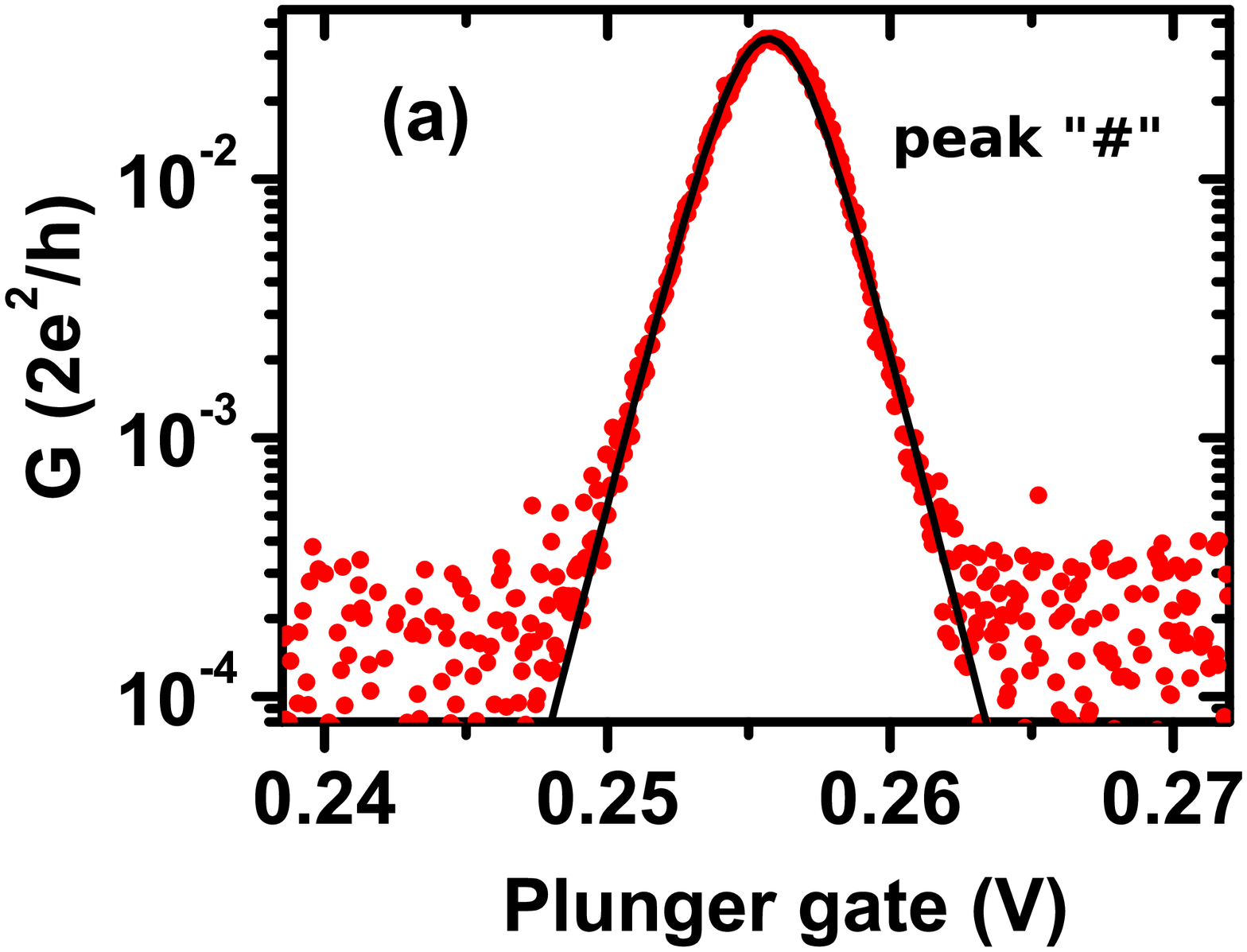}
    \end{minipage}
    \hspace{0.2cm}
    \begin{minipage}{0.47\columnwidth}
    \includegraphics[width=\columnwidth]{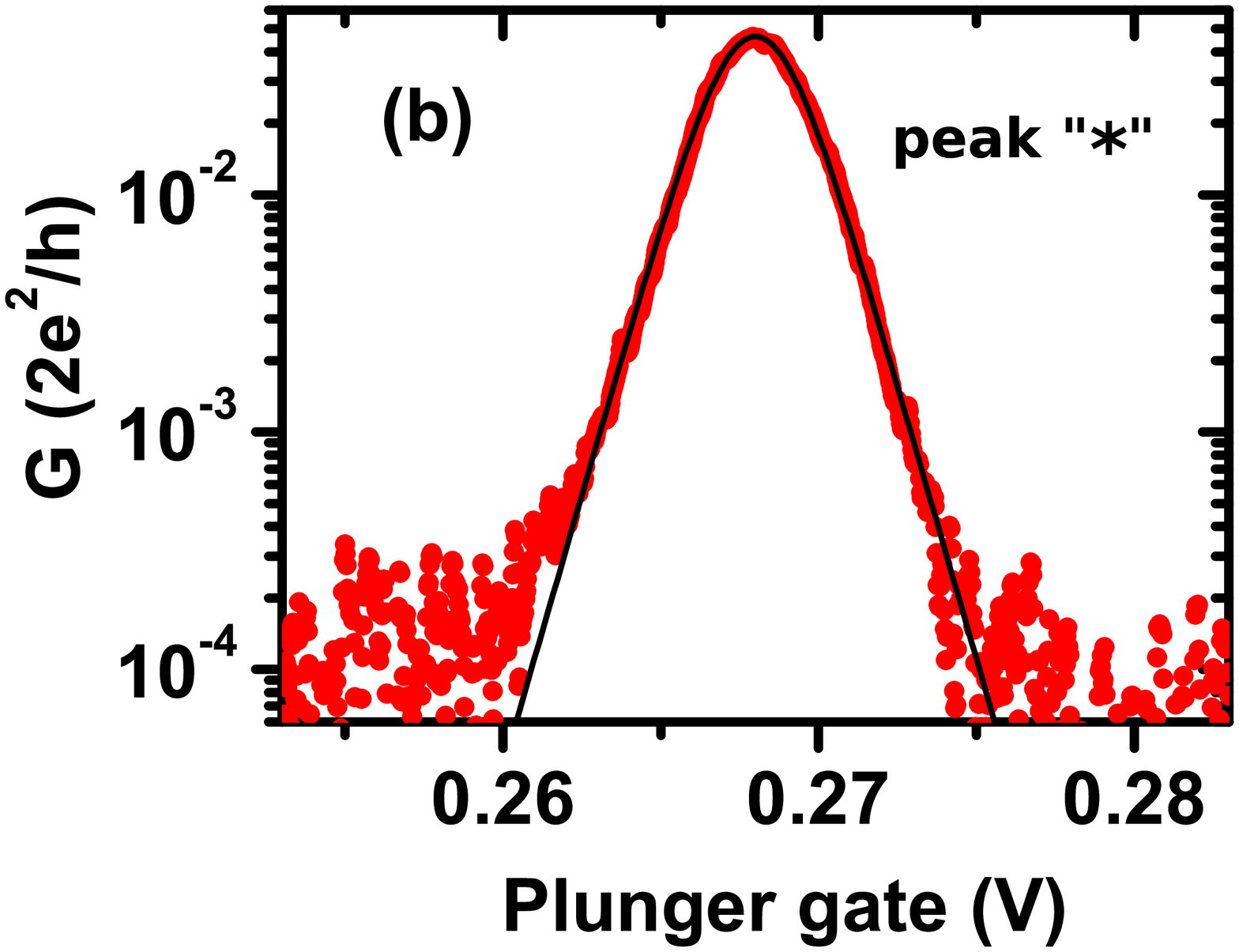}
    \end{minipage}
    \vspace{0.1cm}
    \begin{minipage}{0.47\columnwidth}
    \includegraphics[width=\columnwidth]{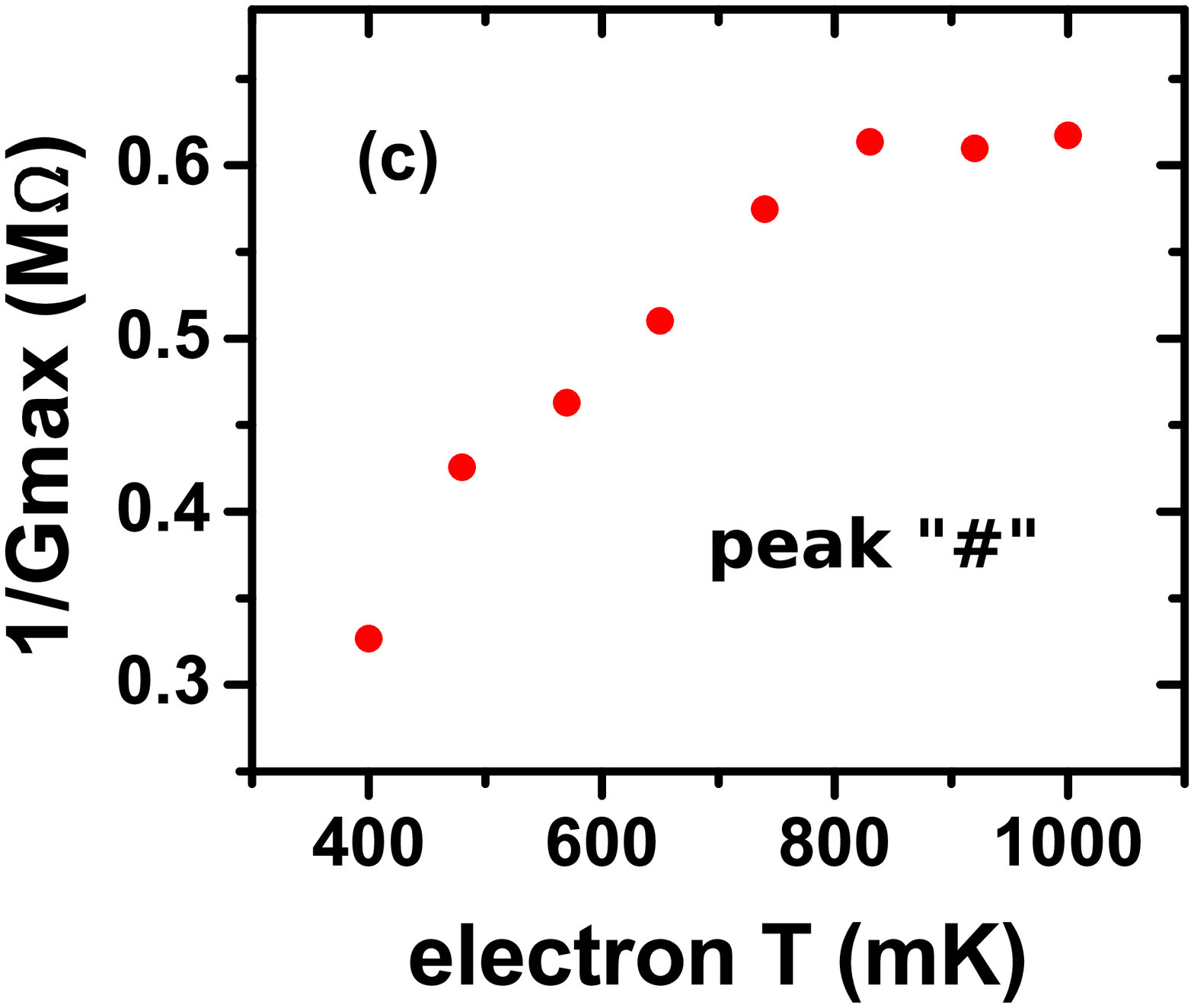}
    \end{minipage}
    \hspace{0.2cm}
    \begin{minipage}{0.47\columnwidth}
    \includegraphics[width=\columnwidth]{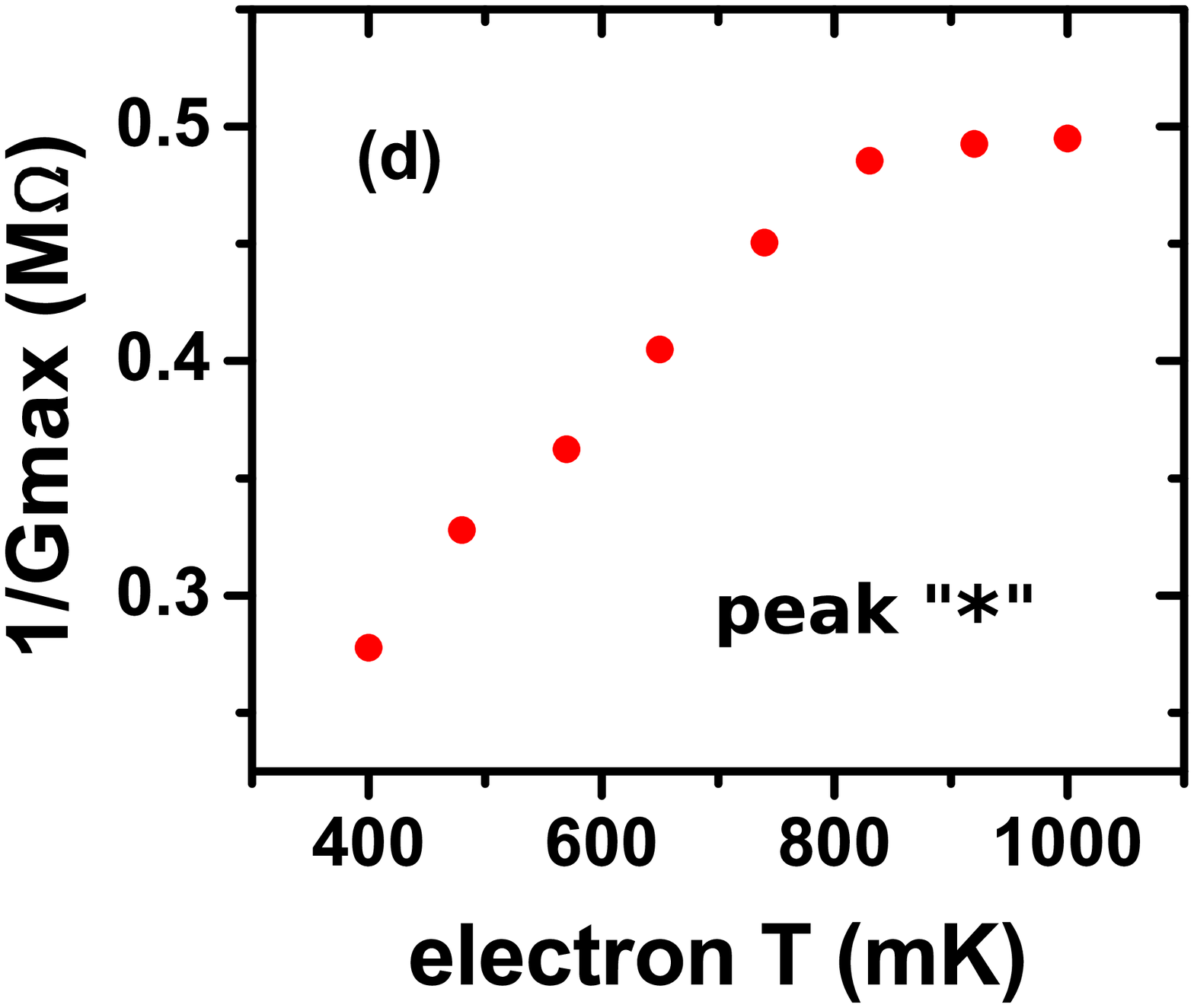}
    \end{minipage}
    \vspace{0.1cm}
    \begin{minipage}{0.46\columnwidth}
    \includegraphics[width=\columnwidth]{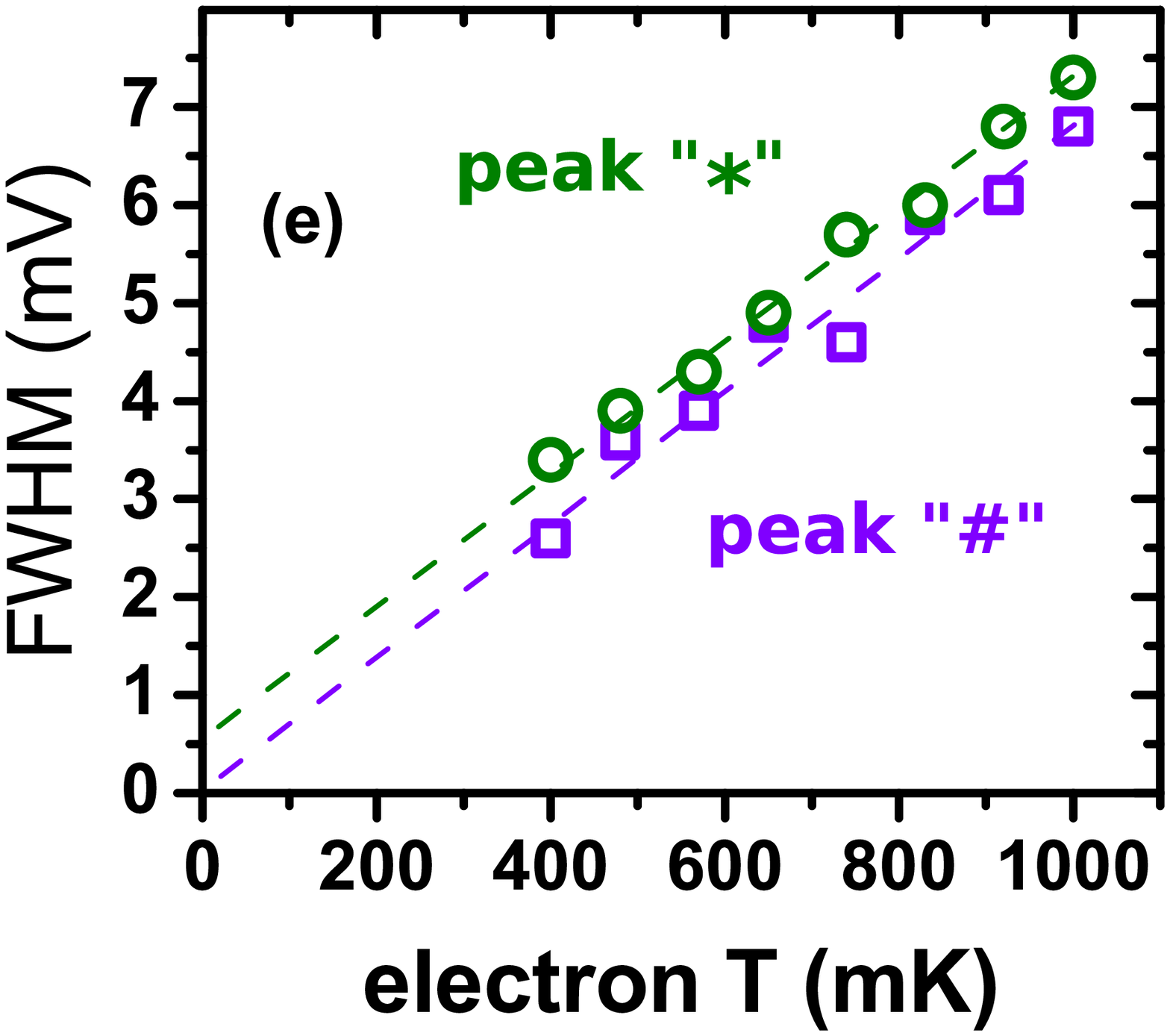}
    \end{minipage}
    \hspace{0.2cm}
    \begin{minipage}{0.47\columnwidth}
    \includegraphics[width=\columnwidth]{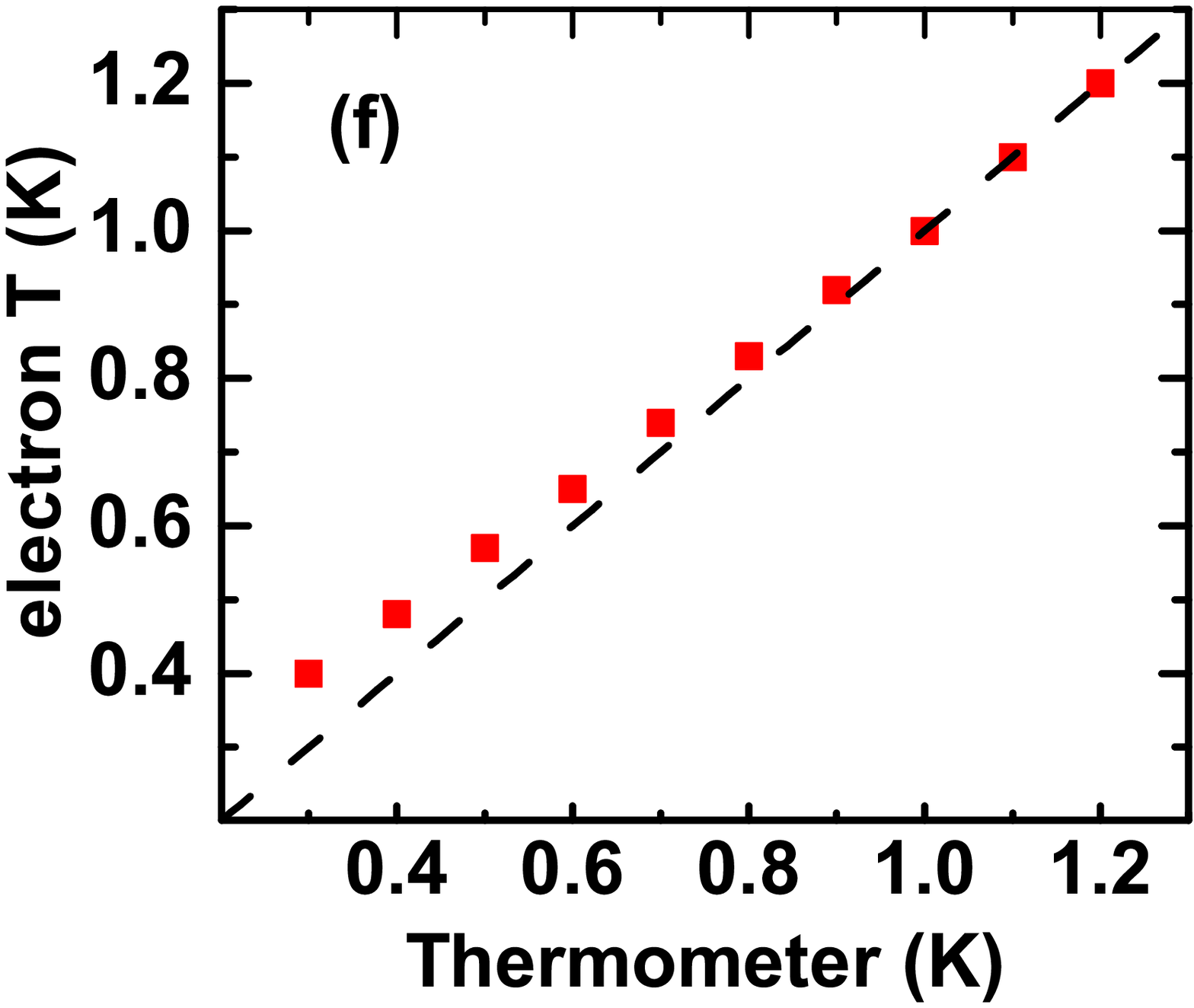}
    \end{minipage}
    \caption{In all panels above, symbols are experimental data from
    the CB peaks marked $\#$ and $\ast$ in Figure \ref{fig2}, solid lines are fits
    to equations, and dashed lines are guides to the eye. Extracted from DC bias
    spectroscopy, the value $\alpha = 0.037$ meV/mV was used in these fits (e-beam
    gates were biased at the same voltages as described in Fig.\,\ref{fig2}).
    (a)-(b) The solid line is a fit to Eqn.\,(\ref{eq:QMdot1}).
    (c)-(d) Below 800\,mK, the inverse of CB peak maximum height (Gmax) is
    proportional to temperature. Paired panels (a)/(c) and (b)/(d) suggest only a
    single energy level is involved in transport for CB peaks $\#$ and $\ast$.
    (e) Full width at half maximum (FWHM) of both peaks against electron temperature.
    (f) Electron temperature deduced from fits of the data to Eq.\,(\ref{eq:QMdot0}).}
    \label{fig4}
\end{figure}

At finite temperature $T$, the conductance $G$ through a quantum dot is determined
by:\cite{Foxman94}
\begin{eqnarray}
G(\mu,T)&=&\frac{e^2}{h}\frac{C_{1}}{kT}~\int_{-\infty}^{+\infty}~
\left[\frac{(h\Gamma)^2}{(E-E_0)^2+(h\Gamma)^2}\right]\nonumber\\
&&\qquad\qquad~~\times~\text{cosh}^{-2} \left[\frac{E-\mu}{2kT}\right]~dE
\label{eq:QMdot0}
\end{eqnarray}
where $C_{1}$ is a constant, $\mu$ the chemical potential, $k$ the
Boltzmann constant, and $\Gamma=(\Gamma_L+\Gamma_R)/2$ is the average of
the tunneling rates through the left and right barriers of the quantum
dot. If the dot is strongly coupled to the leads ($h\Gamma > kT$), the CB
peak lineshape is a Lorentzian [$V_g$\,$>$\,0.35\,V in Fig.\,\ref{fig1}].
In the weak coupling regime, if transport through the dot involves only a
single energy level and $\Gamma$ is very small ($h\Gamma \ll k_BT < \Delta
E < U$), then the lineshape of the CB conductance peak can be approximated
by:\cite{Foxman93,Beenakker91}
\begin{equation}
G(V_g,T)~\approx~\frac{e^2}{h}\frac{C_{2}}{kT}~\text{cosh}^{-2}
\left[\frac{\alpha (V_g-V_0)}{2kT}\right]
\label{eq:QMdot1}
\end{equation}
where $C_{2}$ is a constant, $\alpha$ the lever arm of the plunger gate, $V_g$ the
plunger gate voltage, and $V_0$ the plunger voltage at the center of the CB peak. If
more than one energy level is involved in transport ($h\Gamma \ll \Delta E < k_BT <
U$), then the following approximation can be used:\cite{Foxman94,Beenakker91}
\begin{equation}
G(V_g,T)~\approx~\frac{e^2}{h}\frac{C_{3}}{\Delta E}~\text{cosh}^{-2}
\left[\frac{\alpha (V_g-V_0)}{2.5kT}\right]
\label{eq:QMdot2}
\end{equation}
where $C_{3}$ is a constant and $\Delta E$ is the single-particle energy level
spacing.

Figure \ref{fig4} shows the analysis of the temperature dependence of the
CB peaks labeled ``$\#$'' and ``$\ast$'' in Figure \ref{fig2}. Both peaks
fit the lineshape generated by equation (\ref{eq:QMdot1})
[Figs.\,\ref{fig4}(a) and \ref{fig4}(b)]. This, combined with the linear
temperature dependence of the inverse of the maximum CB peak height
(1/Gmax) in Figs.\,\ref{fig4}(c) and \ref{fig4}(d), strongly suggest
single-level transport occurs at temperatures below 800\,mK. Above
800\,mK, 1/Gmax becomes independent of temperature, signalling a
transition from single-level transport ($kT < \Delta E < U$) to
multi-level transport ($\Delta E < kT < U$) through the quantum dot. The
FWHM of both peaks linearly increases with temperature in both regimes
[Fig.\,\ref{fig4}(e)].

When extrapolated to $T$=0, the FWHM of peak ``$\#$''
[Fig.\,\ref{fig4}(e)] is zero, consistent with the approximation $\Gamma
\rightarrow 0$, used to derive Eqn.\,(\ref{eq:QMdot1}). Similarly, 1/Gmax
goes to zero at $T$=0 [Fig.\,\ref{fig4}(c)]. The story is different for
peak ``$\ast$''. When extrapolated to $T$=0, its FWHM reads the finite
value 0.5\,mV in gate voltage units which, when converted to energy units
(using $\alpha = 37$\,$\mu$eV/mV), gives $h\Gamma = 19\,\mu$eV, a measure
of the lifetime broadening of this energy level. This is roughly half our
base electron temperature. Thus, although Eqn.\,(\ref{eq:QMdot1}) does not
strictly apply, the single-level transport regime is still accessible
(because $h\Gamma < kT$). That $\Gamma \neq 0$ for peak ``$\ast$'' also
explains the presence of a finite offset at $T=0$ for 1/Gmax in
Fig.\,\ref{fig4}(d). Stated otherwise, Gmax does not blow up to infinity
at T=0, but remains finite. This behavior is well
understood.\cite{Foxman94}

In conclusion, we have shown ultra-shallow 2DEGs (within 30\,nm of the surface) in
undoped GaAs/AlGaAs heterostructures that display no parallel conduction, are
gateable, and show no hysteresis upon gate action. The absence of dopants has
improved their mobility by more than one order of magnitude relative to their doped
counterparts with similar 2DEG depths and carrier densities. We have also
demonstrated that these undoped heterostructures can be used to fabricate
single-electron quantum dots defined by surface metal gates, with charging energies
of up to 1.75\,meV and excited state energies of up to 0.5\,meV.

The authors would like to thank S.J. Chorley and C.J.B. Ford for their
help, and acknowledge financial support from Toshiba Research Europe and
the EPSRC.

\bibliographystyle{apsrev4-1}  

\begin{thebibliography}{37}%
\makeatletter
\providecommand \@ifxundefined [1]{%
 \@ifx{#1\undefined}
}%
\providecommand \@ifnum [1]{%
 \ifnum #1\expandafter \@firstoftwo
 \else \expandafter \@secondoftwo
 \fi
}%
\providecommand \@ifx [1]{%
 \ifx #1\expandafter \@firstoftwo
 \else \expandafter \@secondoftwo
 \fi
}%
\providecommand \natexlab [1]{#1}%
\providecommand \enquote  [1]{``#1''}%
\providecommand \bibnamefont  [1]{#1}%
\providecommand \bibfnamefont [1]{#1}%
\providecommand \citenamefont [1]{#1}%
\providecommand \href@noop [0]{\@secondoftwo}%
\providecommand \href [0]{\begingroup \@sanitize@url \@href}%
\providecommand \@href[1]{\@@startlink{#1}\@@href}%
\providecommand \@@href[1]{\endgroup#1\@@endlink}%
\providecommand \@sanitize@url [0]{\catcode `\\12\catcode `\$12\catcode
  `\&12\catcode `\#12\catcode `\^12\catcode `\_12\catcode `\%12\relax}%
\providecommand \@@startlink[1]{}%
\providecommand \@@endlink[0]{}%
\providecommand \url  [0]{\begingroup\@sanitize@url \@url }%
\providecommand \@url [1]{\endgroup\@href {#1}{\urlprefix }}%
\providecommand \urlprefix  [0]{URL }%
\providecommand \Eprint [0]{\href }%
\providecommand \doibase [0]{http://dx.doi.org/}%
\providecommand \selectlanguage [0]{\@gobble}%
\providecommand \bibinfo  [0]{\@secondoftwo}%
\providecommand \bibfield  [0]{\@secondoftwo}%
\providecommand \translation [1]{[#1]}%
\providecommand \BibitemOpen [0]{}%
\providecommand \bibitemStop [0]{}%
\providecommand \bibitemNoStop [0]{.\EOS\space}%
\providecommand \EOS [0]{\spacefactor3000\relax}%
\providecommand \BibitemShut  [1]{\csname bibitem#1\endcsname}%
\let\auto@bib@innerbib\@empty
\bibitem [{\citenamefont {Pioro-Ladriere}\ \emph
    {et~al.}(2005)\citenamefont
  {Pioro-Ladriere}, \citenamefont {Davies}, \citenamefont {Long}, \citenamefont
  {Sachrajda}, \citenamefont {Gaudreau}, \citenamefont {Zawadzki},
  \citenamefont {Lapointe}, \citenamefont {Gupta}, \citenamefont {Wasilewski},\
  and\ \citenamefont {Studenikin}}]{Pioro05-B}%
  \BibitemOpen
  \bibfield  {author} {\bibinfo {author} {\bibfnamefont {M.}~\bibnamefont
  {Pioro-Ladriere}}, \bibinfo {author} {\bibfnamefont {J.~H.}\ \bibnamefont
  {Davies}}, \bibinfo {author} {\bibfnamefont {A.~R.}\ \bibnamefont {Long}},
  \bibinfo {author} {\bibfnamefont {A.~S.}\ \bibnamefont {Sachrajda}}, \bibinfo
  {author} {\bibfnamefont {L.}~\bibnamefont {Gaudreau}}, \bibinfo {author}
  {\bibfnamefont {P.}~\bibnamefont {Zawadzki}}, \bibinfo {author}
  {\bibfnamefont {J.}~\bibnamefont {Lapointe}}, \bibinfo {author}
  {\bibfnamefont {J.}~\bibnamefont {Gupta}}, \bibinfo {author} {\bibfnamefont
  {Z.}~\bibnamefont {Wasilewski}}, \ and\ \bibinfo {author} {\bibfnamefont
  {S.}~\bibnamefont {Studenikin}},\ }\href@noop {} {\bibfield  {journal}
  {\bibinfo  {journal} {Phys. Rev. B}\ }\textbf {\bibinfo {volume} {72}},\
  \bibinfo {pages} {115331} (\bibinfo {year} {2005})}\BibitemShut {NoStop}%
\bibitem [{\citenamefont {Buizert}\ \emph {et~al.}(2008)\citenamefont
  {Buizert}, \citenamefont {Koppens}, \citenamefont {Pioro-Ladriere},
  \citenamefont {Tranitz}, \citenamefont {Vink}, \citenamefont {Tarucha},
  \citenamefont {Wegscheider},\ and\ \citenamefont {Vandersypen}}]{Buizert08}%
  \BibitemOpen
  \bibfield  {author} {\bibinfo {author} {\bibfnamefont {C.}~\bibnamefont
  {Buizert}}, \bibinfo {author} {\bibfnamefont {F.~H.~L.}\ \bibnamefont
  {Koppens}}, \bibinfo {author} {\bibfnamefont {M.}~\bibnamefont
  {Pioro-Ladriere}}, \bibinfo {author} {\bibfnamefont {H.~P.}\ \bibnamefont
  {Tranitz}}, \bibinfo {author} {\bibfnamefont {I.~T.}\ \bibnamefont {Vink}},
  \bibinfo {author} {\bibfnamefont {S.}~\bibnamefont {Tarucha}}, \bibinfo
  {author} {\bibfnamefont {W.}~\bibnamefont {Wegscheider}}, \ and\ \bibinfo
  {author} {\bibfnamefont {L.~M.~K.}\ \bibnamefont {Vandersypen}},\ }\href@noop
  {} {\bibfield  {journal} {\bibinfo  {journal} {Phys. Rev. Lett.}\ }\textbf
  {\bibinfo {volume} {101}},\ \bibinfo {pages} {226603} (\bibinfo {year}
  {2008})}\BibitemShut {NoStop}%
\bibitem [{\citenamefont {Fujiwara}\ \emph {et~al.}(2002)\citenamefont
  {Fujiwara}, \citenamefont {Sasaki},\ and\ \citenamefont
  {Akiba}}]{Fujiwara02}%
  \BibitemOpen
  \bibfield  {author} {\bibinfo {author} {\bibfnamefont {M.}~\bibnamefont
  {Fujiwara}}, \bibinfo {author} {\bibfnamefont {M.}~\bibnamefont {Sasaki}}, \
  and\ \bibinfo {author} {\bibfnamefont {M.}~\bibnamefont {Akiba}},\
  }\href@noop {} {\bibfield  {journal} {\bibinfo  {journal} {Appl. Phys.
  Lett.}\ }\textbf {\bibinfo {volume} {80}},\ \bibinfo {pages} {1844} (\bibinfo
  {year} {2002})}\BibitemShut {NoStop}%
\bibitem [{\citenamefont {Kopnov}\ \emph {et~al.}(2010)\citenamefont
    {Kopnov},
  \citenamefont {Umansky}, \citenamefont {Cohen}, \citenamefont {Shahar},\ and\
  \citenamefont {Naaman}}]{Kopnov10}%
  \BibitemOpen
  \bibfield  {author} {\bibinfo {author} {\bibfnamefont {G.}~\bibnamefont
  {Kopnov}}, \bibinfo {author} {\bibfnamefont {V.~Y.}\ \bibnamefont {Umansky}},
  \bibinfo {author} {\bibfnamefont {H.}~\bibnamefont {Cohen}}, \bibinfo
  {author} {\bibfnamefont {D.}~\bibnamefont {Shahar}}, \ and\ \bibinfo {author}
  {\bibfnamefont {R.}~\bibnamefont {Naaman}},\ }\href@noop {} {\bibfield
  {journal} {\bibinfo  {journal} {Phys. Rev. B}\ }\textbf {\bibinfo {volume}
  {81}},\ \bibinfo {pages} {045316} (\bibinfo {year} {2010})}\BibitemShut
  {NoStop}%
\bibitem [{\citenamefont {Laroche}\ \emph {et~al.}(2010)\citenamefont
  {Laroche}, \citenamefont {{Das~Sarma}}, \citenamefont {Gervais},
  \citenamefont {Lilly},\ and\ \citenamefont {Reno}}]{Laroche10}%
  \BibitemOpen
  \bibfield  {author} {\bibinfo {author} {\bibfnamefont {D.}~\bibnamefont
  {Laroche}}, \bibinfo {author} {\bibfnamefont {S.}~\bibnamefont
  {{Das~Sarma}}}, \bibinfo {author} {\bibfnamefont {G.}~\bibnamefont
  {Gervais}}, \bibinfo {author} {\bibfnamefont {M.~P.}\ \bibnamefont {Lilly}},
  \ and\ \bibinfo {author} {\bibfnamefont {J.~L.}\ \bibnamefont {Reno}},\
  }\href@noop {} {\bibfield  {journal} {\bibinfo  {journal} {Appl. Phys.
  Lett.}\ }\textbf {\bibinfo {volume} {96}},\ \bibinfo {pages} {162112}
  (\bibinfo {year} {2010})}\BibitemShut {NoStop}%
\bibitem [{\citenamefont {Giesbers}\ \emph {et~al.}(2010)\citenamefont
  {Giesbers}, \citenamefont {Zeitler}, \citenamefont {Katsnelson},
  \citenamefont {Reuter}, \citenamefont {Wieck}, \citenamefont {Biasiol},
  \citenamefont {Sorba},\ and\ \citenamefont {Maan}}]{Giesbers10}%
  \BibitemOpen
  \bibfield  {author} {\bibinfo {author} {\bibfnamefont {A.~J.~M.}\
  \bibnamefont {Giesbers}}, \bibinfo {author} {\bibfnamefont {U.}~\bibnamefont
  {Zeitler}}, \bibinfo {author} {\bibfnamefont {M.~I.}\ \bibnamefont
  {Katsnelson}}, \bibinfo {author} {\bibfnamefont {D.}~\bibnamefont {Reuter}},
  \bibinfo {author} {\bibfnamefont {A.~D.}\ \bibnamefont {Wieck}}, \bibinfo
  {author} {\bibfnamefont {G.}~\bibnamefont {Biasiol}}, \bibinfo {author}
  {\bibfnamefont {L.}~\bibnamefont {Sorba}}, \ and\ \bibinfo {author}
  {\bibfnamefont {J.~C.}\ \bibnamefont {Maan}},\ }\href@noop {} {\bibfield
  {journal} {\bibinfo  {journal} {Nature Phys.}\ }\textbf {\bibinfo {volume} {6}},\
  \bibinfo {pages} {173} (\bibinfo {year} {2010})}\BibitemShut {NoStop}%
\bibitem [{\citenamefont {Rossler}\ \emph
  {et~al.}(2010{\natexlab{a}})\citenamefont {Rossler}, \citenamefont
  {K{\"{u}}ng}, \citenamefont {Dr{\"{u}}scher}, \citenamefont {Choi},
  \citenamefont {Ihn}, \citenamefont {Ensslin},\ and\ \citenamefont
  {Beck}}]{Rossler10-B}%
  \BibitemOpen
  \bibfield  {author} {\bibinfo {author} {\bibfnamefont {C.}~\bibnamefont
  {Rossler}}, \bibinfo {author} {\bibfnamefont {B.}~\bibnamefont {K{\"{u}}ng}},
  \bibinfo {author} {\bibfnamefont {S.}~\bibnamefont {Dr{\"{u}}scher}},
  \bibinfo {author} {\bibfnamefont {T.}~\bibnamefont {Choi}}, \bibinfo {author}
  {\bibfnamefont {T.}~\bibnamefont {Ihn}}, \bibinfo {author} {\bibfnamefont
  {K.}~\bibnamefont {Ensslin}}, \ and\ \bibinfo {author} {\bibfnamefont
  {M.}~\bibnamefont {Beck}},\ }\href@noop {} {\bibfield  {journal} {\bibinfo
  {journal} {Appl. Phys. Lett.}\ }\textbf {\bibinfo {volume} {97}},\ \bibinfo
  {pages} {152109} (\bibinfo {year} {2010}{\natexlab{a}})}\BibitemShut
  {NoStop}%
\bibitem [{\citenamefont {Nemutudi}\ \emph {et~al.}(2005)\citenamefont
  {Nemutudi}, \citenamefont {Liang}, \citenamefont {Murphy}, \citenamefont
  {Beere}, \citenamefont {Smith}, \citenamefont {Ritchie}, \citenamefont
  {Pepper},\ and\ \citenamefont {Jones}}]{Nemutudi05}%
  \BibitemOpen
  \bibfield  {author} {\bibinfo {author} {\bibfnamefont {R.}~\bibnamefont
  {Nemutudi}}, \bibinfo {author} {\bibfnamefont {C.-T.}\ \bibnamefont {Liang}},
  \bibinfo {author} {\bibfnamefont {M.~J.}\ \bibnamefont {Murphy}}, \bibinfo
  {author} {\bibfnamefont {H.~E.}\ \bibnamefont {Beere}}, \bibinfo {author}
  {\bibfnamefont {C.~G.}\ \bibnamefont {Smith}}, \bibinfo {author}
  {\bibfnamefont {D.~A.}\ \bibnamefont {Ritchie}}, \bibinfo {author}
  {\bibfnamefont {M.}~\bibnamefont {Pepper}}, \ and\ \bibinfo {author}
  {\bibfnamefont {G.~A.~C.}\ \bibnamefont {Jones}},\ }\href@noop {} {\bibfield
  {journal} {\bibinfo  {journal} {Microelectronics Journal}\ }\textbf {\bibinfo
  {volume} {36}},\ \bibinfo {pages} {425} (\bibinfo {year} {2005})}\BibitemShut
  {NoStop}%
\bibitem [{\citenamefont {Gildemeister}\ \emph {et~al.}(2007)\citenamefont
  {Gildemeister}, \citenamefont {Ihn}, \citenamefont {Schleser}, \citenamefont
  {Ensslin}, \citenamefont {Driscoll},\ and\ \citenamefont
  {Gossard}}]{Gildmeister07}%
  \BibitemOpen
  \bibfield  {author} {\bibinfo {author} {\bibfnamefont {A.~E.}\ \bibnamefont
  {Gildemeister}}, \bibinfo {author} {\bibfnamefont {T.}~\bibnamefont {Ihn}},
  \bibinfo {author} {\bibfnamefont {R.}~\bibnamefont {Schleser}}, \bibinfo
  {author} {\bibfnamefont {K.}~\bibnamefont {Ensslin}}, \bibinfo {author}
  {\bibfnamefont {D.~C.}\ \bibnamefont {Driscoll}}, \ and\ \bibinfo {author}
  {\bibfnamefont {A.~C.}\ \bibnamefont {Gossard}},\ }\href@noop {} {\bibfield
  {journal} {\bibinfo  {journal} {J. Appl. Phys.}\ }\textbf {\bibinfo {volume}
  {102}},\ \bibinfo {pages} {083703} (\bibinfo {year} {2007})}\BibitemShut
  {NoStop}%
\bibitem [{\citenamefont {Nakamura}\ \emph {et~al.}(2010)\citenamefont
  {Nakamura}, \citenamefont {Yamauchi}, \citenamefont {Hashisaka},
  \citenamefont {Chida}, \citenamefont {Kobayashi}, \citenamefont {Ono},
  \citenamefont {Leturcq}, \citenamefont {Ensslin}, \citenamefont {Saito},
  \citenamefont {Utsumi},\ and\ \citenamefont {Gossard}}]{Nakamura10}%
  \BibitemOpen
  \bibfield  {author} {\bibinfo {author} {\bibfnamefont {S.}~\bibnamefont
  {Nakamura}}, \bibinfo {author} {\bibfnamefont {Y.}~\bibnamefont {Yamauchi}},
  \bibinfo {author} {\bibfnamefont {M.}~\bibnamefont {Hashisaka}}, \bibinfo
  {author} {\bibfnamefont {K.}~\bibnamefont {Chida}}, \bibinfo {author}
  {\bibfnamefont {K.}~\bibnamefont {Kobayashi}}, \bibinfo {author}
  {\bibfnamefont {T.}~\bibnamefont {Ono}}, \bibinfo {author} {\bibfnamefont
  {R.}~\bibnamefont {Leturcq}}, \bibinfo {author} {\bibfnamefont
  {K.}~\bibnamefont {Ensslin}}, \bibinfo {author} {\bibfnamefont
  {K.}~\bibnamefont {Saito}}, \bibinfo {author} {\bibfnamefont
  {Y.}~\bibnamefont {Utsumi}}, \ and\ \bibinfo {author} {\bibfnamefont {A.~C.}\
  \bibnamefont {Gossard}},\ }\href@noop {} {\bibfield  {journal} {\bibinfo
  {journal} {Phys. Rev. Lett.}\ }\textbf {\bibinfo {volume} {104}},\ \bibinfo
  {pages} {080602} (\bibinfo {year} {2010})}\BibitemShut {NoStop}%
\bibitem [{\citenamefont {M{\"{u}}hle}\ \emph {et~al.}(2008)\citenamefont
  {M{\"{u}}hle}, \citenamefont {Wegscheider},\ and\ \citenamefont
  {Haug}}]{Muhle08}%
  \BibitemOpen
  \bibfield  {author} {\bibinfo {author} {\bibfnamefont {A.}~\bibnamefont
  {M{\"{u}}hle}}, \bibinfo {author} {\bibfnamefont {W.}~\bibnamefont
  {Wegscheider}}, \ and\ \bibinfo {author} {\bibfnamefont {R.~J.}\ \bibnamefont
  {Haug}},\ }\href@noop {} {\bibfield  {journal} {\bibinfo  {journal} {Appl.
  Phys. Lett.}\ }\textbf {\bibinfo {volume} {92}},\ \bibinfo {pages} {013126}
  (\bibinfo {year} {2008})}\BibitemShut {NoStop}%
\bibitem [{\citenamefont {Fricke}\ \emph {et~al.}(2007)\citenamefont
    {Fricke},
  \citenamefont {Hohls}, \citenamefont {Wegscheider},\ and\ \citenamefont
  {Haug}}]{Fricke07}%
  \BibitemOpen
  \bibfield  {author} {\bibinfo {author} {\bibfnamefont {C.}~\bibnamefont
  {Fricke}}, \bibinfo {author} {\bibfnamefont {F.}~\bibnamefont {Hohls}},
  \bibinfo {author} {\bibfnamefont {W.}~\bibnamefont {Wegscheider}}, \ and\
  \bibinfo {author} {\bibfnamefont {R.~J.}\ \bibnamefont {Haug}},\ }\href@noop
  {} {\bibfield  {journal} {\bibinfo  {journal} {Phys. Rev. B}\ }\textbf
  {\bibinfo {volume} {76}},\ \bibinfo {pages} {155307} (\bibinfo {year}
  {2007})}\BibitemShut {NoStop}%
\bibitem [{\citenamefont {Nemutudi}\ \emph {et~al.}(2004)\citenamefont
  {Nemutudi}, \citenamefont {Kataoka}, \citenamefont {Ford}, \citenamefont
  {Appleyard}, \citenamefont {Pepper}, \citenamefont {Ritchie},\ and\
  \citenamefont {Jones}}]{Nemutudi04}%
  \BibitemOpen
  \bibfield  {author} {\bibinfo {author} {\bibfnamefont {R.}~\bibnamefont
  {Nemutudi}}, \bibinfo {author} {\bibfnamefont {M.}~\bibnamefont {Kataoka}},
  \bibinfo {author} {\bibfnamefont {C.~J.~B.}\ \bibnamefont {Ford}}, \bibinfo
  {author} {\bibfnamefont {N.~J.}\ \bibnamefont {Appleyard}}, \bibinfo {author}
  {\bibfnamefont {M.}~\bibnamefont {Pepper}}, \bibinfo {author} {\bibfnamefont
  {D.~A.}\ \bibnamefont {Ritchie}}, \ and\ \bibinfo {author} {\bibfnamefont
  {G.~A.~C.}\ \bibnamefont {Jones}},\ }\href@noop {} {\bibfield  {journal}
  {\bibinfo  {journal} {J. Appl. Phys.}\ }\textbf {\bibinfo {volume} {95}},\
  \bibinfo {pages} {2557} (\bibinfo {year} {2004})}\BibitemShut {NoStop}%
\bibitem [{\citenamefont {Goldhaber-Gordon}\ \emph
    {et~al.}(1998)\citenamefont
  {Goldhaber-Gordon}, \citenamefont {Shtrikman}, \citenamefont {Mahalu},
  \citenamefont {Abusch-Magder}, \citenamefont {Meirav},\ and\ \citenamefont
  {Kastner}}]{Goldhaber98-A}%
  \BibitemOpen
  \bibfield  {author} {\bibinfo {author} {\bibfnamefont {D.}~\bibnamefont
  {Goldhaber-Gordon}}, \bibinfo {author} {\bibfnamefont {H.}~\bibnamefont
  {Shtrikman}}, \bibinfo {author} {\bibfnamefont {D.}~\bibnamefont {Mahalu}},
  \bibinfo {author} {\bibfnamefont {D.}~\bibnamefont {Abusch-Magder}}, \bibinfo
  {author} {\bibfnamefont {U.}~\bibnamefont {Meirav}}, \ and\ \bibinfo {author}
  {\bibfnamefont {M.~A.}\ \bibnamefont {Kastner}},\ }\href@noop {} {\bibfield
  {journal} {\bibinfo  {journal} {Nature}\ }\textbf {\bibinfo {volume} {391}},\
  \bibinfo {pages} {156} (\bibinfo {year} {1998})}\BibitemShut {NoStop}%
\bibitem [{\citenamefont {Miller}\ \emph {et~al.}(2007)\citenamefont
    {Miller},
  \citenamefont {Radu}, \citenamefont {Zumbuhl}, \citenamefont {Levenson-Falk},
  \citenamefont {Kastner}, \citenamefont {Marcus}, \citenamefont {Pfeiffer},\
  and\ \citenamefont {West}}]{Miller07}%
  \BibitemOpen
  \bibfield  {author} {\bibinfo {author} {\bibfnamefont {J.~B.}\ \bibnamefont
  {Miller}}, \bibinfo {author} {\bibfnamefont {I.~P.}\ \bibnamefont {Radu}},
  \bibinfo {author} {\bibfnamefont {D.~M.}\ \bibnamefont {Zumbuhl}}, \bibinfo
  {author} {\bibfnamefont {E.~M.}\ \bibnamefont {Levenson-Falk}}, \bibinfo
  {author} {\bibfnamefont {M.~A.}\ \bibnamefont {Kastner}}, \bibinfo {author}
  {\bibfnamefont {C.~M.}\ \bibnamefont {Marcus}}, \bibinfo {author}
  {\bibfnamefont {L.~N.}\ \bibnamefont {Pfeiffer}}, \ and\ \bibinfo {author}
  {\bibfnamefont {K.~W.}\ \bibnamefont {West}},\ }\href@noop {} {\bibfield
  {journal} {\bibinfo  {journal} {Nature Phys.}\ }\textbf {\bibinfo {volume}
  {3}},\ \bibinfo {pages} {561} (\bibinfo {year} {2007})}\BibitemShut {NoStop}%
\bibitem [{\citenamefont {Dolev}\ \emph {et~al.}(2008)\citenamefont
    {Dolev},
  \citenamefont {Heiblum}, \citenamefont {Umansky}, \citenamefont {Stern},\
  and\ \citenamefont {Mahalu}}]{Dolev08}%
  \BibitemOpen
  \bibfield  {author} {\bibinfo {author} {\bibfnamefont {M.}~\bibnamefont
  {Dolev}}, \bibinfo {author} {\bibfnamefont {M.}~\bibnamefont {Heiblum}},
  \bibinfo {author} {\bibfnamefont {V.}~\bibnamefont {Umansky}}, \bibinfo
  {author} {\bibfnamefont {A.}~\bibnamefont {Stern}}, \ and\ \bibinfo {author}
  {\bibfnamefont {D.}~\bibnamefont {Mahalu}},\ }\href@noop {} {\bibfield
  {journal} {\bibinfo  {journal} {Nature}\ }\textbf {\bibinfo {volume} {452}},\
  \bibinfo {pages} {829} (\bibinfo {year} {2008})}\BibitemShut {NoStop}%
\bibitem [{\citenamefont {Rossler}\ \emph
  {et~al.}(2010{\natexlab{b}})\citenamefont {Rossler}, \citenamefont {Feil},
  \citenamefont {Mensch}, \citenamefont {Ihn}, \citenamefont {Ensslin},
  \citenamefont {Schuh},\ and\ \citenamefont {Wegscheider}}]{Rossler10-A}%
  \BibitemOpen
  \bibfield  {author} {\bibinfo {author} {\bibfnamefont {C.}~\bibnamefont
  {Rossler}}, \bibinfo {author} {\bibfnamefont {T.}~\bibnamefont {Feil}},
  \bibinfo {author} {\bibfnamefont {P.}~\bibnamefont {Mensch}}, \bibinfo
  {author} {\bibfnamefont {T.}~\bibnamefont {Ihn}}, \bibinfo {author}
  {\bibfnamefont {K.}~\bibnamefont {Ensslin}}, \bibinfo {author} {\bibfnamefont
  {D.}~\bibnamefont {Schuh}}, \ and\ \bibinfo {author} {\bibfnamefont
  {W.}~\bibnamefont {Wegscheider}},\ }\href@noop {} {\bibfield  {journal}
  {\bibinfo  {journal} {New J. Phys.}\ }\textbf {\bibinfo {volume} {12}},\
  \bibinfo {pages} {043007} (\bibinfo {year} {2010}{\natexlab{b}})}\BibitemShut
  {NoStop}%
\bibitem [{\citenamefont {Kane}\ \emph {et~al.}(1993)\citenamefont {Kane},
  \citenamefont {Pfeiffer}, \citenamefont {West},\ and\ \citenamefont
  {Harnett}}]{Kane93}%
  \BibitemOpen
  \bibfield  {author} {\bibinfo {author} {\bibfnamefont {B.~E.}\ \bibnamefont
  {Kane}}, \bibinfo {author} {\bibfnamefont {L.~N.}\ \bibnamefont {Pfeiffer}},
  \bibinfo {author} {\bibfnamefont {K.~W.}\ \bibnamefont {West}}, \ and\
  \bibinfo {author} {\bibfnamefont {C.~K.}\ \bibnamefont {Harnett}},\
  }\href@noop {} {\bibfield  {journal} {\bibinfo  {journal} {Appl. Phys.
  Lett.}\ }\textbf {\bibinfo {volume} {63}},\ \bibinfo {pages} {2132} (\bibinfo
  {year} {1993})}\BibitemShut {NoStop}%
\bibitem [{\citenamefont {Kawaharazuka}\ \emph {et~al.}(2001)\citenamefont
  {Kawaharazuka}, \citenamefont {Saku}, \citenamefont {Kikuchi}, \citenamefont
  {Horikoshi},\ and\ \citenamefont {Hirayama}}]{Kawaharazuka01}%
  \BibitemOpen
  \bibfield  {author} {\bibinfo {author} {\bibfnamefont {A.}~\bibnamefont
  {Kawaharazuka}}, \bibinfo {author} {\bibfnamefont {T.}~\bibnamefont {Saku}},
  \bibinfo {author} {\bibfnamefont {C.~A.}\ \bibnamefont {Kikuchi}}, \bibinfo
  {author} {\bibfnamefont {Y.}~\bibnamefont {Horikoshi}}, \ and\ \bibinfo
  {author} {\bibfnamefont {Y.}~\bibnamefont {Hirayama}},\ }\href@noop {}
  {\bibfield  {journal} {\bibinfo  {journal} {Phys. Rev. B}\ }\textbf {\bibinfo
  {volume} {63}},\ \bibinfo {pages} {245309} (\bibinfo {year}
  {2001})}\BibitemShut {NoStop}%
\bibitem [{\citenamefont {Reilly}\ \emph {et~al.}(2002)\citenamefont
    {Reilly},
  \citenamefont {Buehler}, \citenamefont {OBrien}, \citenamefont {Hamilton},
  \citenamefont {Dzurak}, \citenamefont {Clark}, \citenamefont {Kane},
  \citenamefont {Pfeiffer},\ and\ \citenamefont {West}}]{Reilly02}%
  \BibitemOpen
  \bibfield  {author} {\bibinfo {author} {\bibfnamefont {D.~J.}\ \bibnamefont
  {Reilly}}, \bibinfo {author} {\bibfnamefont {T.~M.}\ \bibnamefont {Buehler}},
  \bibinfo {author} {\bibfnamefont {J.~L.}\ \bibnamefont {OBrien}}, \bibinfo
  {author} {\bibfnamefont {A.~R.}\ \bibnamefont {Hamilton}}, \bibinfo {author}
  {\bibfnamefont {A.~S.}\ \bibnamefont {Dzurak}}, \bibinfo {author}
  {\bibfnamefont {R.~G.}\ \bibnamefont {Clark}}, \bibinfo {author}
  {\bibfnamefont {B.~E.}\ \bibnamefont {Kane}}, \bibinfo {author}
  {\bibfnamefont {L.~N.}\ \bibnamefont {Pfeiffer}}, \ and\ \bibinfo {author}
  {\bibfnamefont {K.~W.}\ \bibnamefont {West}},\ }\href@noop {} {\bibfield
  {journal} {\bibinfo  {journal} {Phys. Rev. Lett.}\ }\textbf {\bibinfo
  {volume} {89}},\ \bibinfo {pages} {246801} (\bibinfo {year}
  {2002})}\BibitemShut {NoStop}%
\bibitem [{\citenamefont {Noh}\ \emph {et~al.}(2003)\citenamefont {Noh},
  \citenamefont {Lilly}, \citenamefont {Tsui}, \citenamefont {Simmons},
  \citenamefont {Hwang}, \citenamefont {{Das~Sarma}}, \citenamefont
  {Pfeiffer},\ and\ \citenamefont {West}}]{Noh03-A}%
  \BibitemOpen
  \bibfield  {author} {\bibinfo {author} {\bibfnamefont {H.}~\bibnamefont
  {Noh}}, \bibinfo {author} {\bibfnamefont {M.~P.}\ \bibnamefont {Lilly}},
  \bibinfo {author} {\bibfnamefont {D.~C.}\ \bibnamefont {Tsui}}, \bibinfo
  {author} {\bibfnamefont {J.~A.}\ \bibnamefont {Simmons}}, \bibinfo {author}
  {\bibfnamefont {E.~H.}\ \bibnamefont {Hwang}}, \bibinfo {author}
  {\bibfnamefont {S.}~\bibnamefont {{Das~Sarma}}}, \bibinfo {author}
  {\bibfnamefont {L.~N.}\ \bibnamefont {Pfeiffer}}, \ and\ \bibinfo {author}
  {\bibfnamefont {K.~W.}\ \bibnamefont {West}},\ }\href@noop {} {\bibfield
  {journal} {\bibinfo  {journal} {Phys. Rev. B}\ }\textbf {\bibinfo {volume}
  {68}},\ \bibinfo {pages} {165308} (\bibinfo {year} {2003})}\BibitemShut
  {NoStop}%
\bibitem [{\citenamefont {Lilly}\ \emph {et~al.}(2003)\citenamefont
    {Lilly},
  \citenamefont {Reno}, \citenamefont {Simmons}, \citenamefont {Spielman},
  \citenamefont {Eisenstein}, \citenamefont {Pfeiffer},\ and\ \citenamefont
  {West}}]{Lilly03}%
  \BibitemOpen
  \bibfield  {author} {\bibinfo {author} {\bibfnamefont {M.~P.}\ \bibnamefont
  {Lilly}}, \bibinfo {author} {\bibfnamefont {J.~L.}\ \bibnamefont {Reno}},
  \bibinfo {author} {\bibfnamefont {J.~A.}\ \bibnamefont {Simmons}}, \bibinfo
  {author} {\bibfnamefont {I.~B.}\ \bibnamefont {Spielman}}, \bibinfo {author}
  {\bibfnamefont {J.~P.}\ \bibnamefont {Eisenstein}}, \bibinfo {author}
  {\bibfnamefont {L.~N.}\ \bibnamefont {Pfeiffer}}, \ and\ \bibinfo {author}
  {\bibfnamefont {K.~W.}\ \bibnamefont {West}},\ }\href@noop {} {\bibfield
  {journal} {\bibinfo  {journal} {Phys. Rev. Lett.}\ }\textbf {\bibinfo
  {volume} {90}},\ \bibinfo {pages} {056806} (\bibinfo {year}
  {2003})}\BibitemShut {NoStop}%
\bibitem [{\citenamefont {Herfort}\ and\ \citenamefont
  {Hirayama}(1996)}]{Herfort96}%
  \BibitemOpen
  \bibfield  {author} {\bibinfo {author} {\bibfnamefont {J.}~\bibnamefont
  {Herfort}}\ and\ \bibinfo {author} {\bibfnamefont {Y.}~\bibnamefont
  {Hirayama}},\ }\href@noop {} {\bibfield  {journal} {\bibinfo  {journal}
  {Appl. Phys. Lett.}\ }\textbf {\bibinfo {volume} {69}},\ \bibinfo {pages}
  {3360} (\bibinfo {year} {1996})}\BibitemShut {NoStop}%
\bibitem [{\citenamefont {Harrell}\ \emph {et~al.}(1999)\citenamefont
  {Harrell}, \citenamefont {Pyshkin}, \citenamefont {Simmons}, \citenamefont
  {Ritchie}, \citenamefont {Ford}, \citenamefont {Jones},\ and\ \citenamefont
  {Pepper}}]{Harrell99}%
  \BibitemOpen
  \bibfield  {author} {\bibinfo {author} {\bibfnamefont {R.~H.}\ \bibnamefont
  {Harrell}}, \bibinfo {author} {\bibfnamefont {K.~S.}\ \bibnamefont
  {Pyshkin}}, \bibinfo {author} {\bibfnamefont {M.~Y.}\ \bibnamefont
  {Simmons}}, \bibinfo {author} {\bibfnamefont {D.~A.}\ \bibnamefont
  {Ritchie}}, \bibinfo {author} {\bibfnamefont {C.~J.~B.}\ \bibnamefont
  {Ford}}, \bibinfo {author} {\bibfnamefont {G.~A.~C.}\ \bibnamefont {Jones}},
  \ and\ \bibinfo {author} {\bibfnamefont {M.}~\bibnamefont {Pepper}},\
  }\href@noop {} {\bibfield  {journal} {\bibinfo  {journal} {Appl. Phys.
  Lett.}\ }\textbf {\bibinfo {volume} {74}},\ \bibinfo {pages} {2328} (\bibinfo
  {year} {1999})}\BibitemShut {NoStop}%
\bibitem [{\citenamefont {Willett}\ \emph {et~al.}(2006)\citenamefont
  {Willett}, \citenamefont {Pfeiffer},\ and\ \citenamefont {West}}]{Willett06}%
  \BibitemOpen
  \bibfield  {author} {\bibinfo {author} {\bibfnamefont {R.~L.}\ \bibnamefont
  {Willett}}, \bibinfo {author} {\bibfnamefont {L.~N.}\ \bibnamefont
  {Pfeiffer}}, \ and\ \bibinfo {author} {\bibfnamefont {K.~W.}\ \bibnamefont
  {West}},\ }\href@noop {} {\bibfield  {journal} {\bibinfo  {journal} {Appl.
  Phys. Lett.}\ }\textbf {\bibinfo {volume} {89}},\ \bibinfo {pages} {242107}
  (\bibinfo {year} {2006})}\BibitemShut {NoStop}%
\bibitem [{Note1()}]{Note1}%
  \BibitemOpen
  \bibinfo {note} {The term HIGFET has been used in the literature to refer to
  both SISFET and/or MISFET geometries. By MISFET, we mean the gate is a metal
  (e.g. Au) and the insulator is some dielectric. By SISFET, we mean the gate
  is a highly-doped semiconductor (metallic regime) and the insulator is also a
  semiconductor.}\BibitemShut {Stop}%
\bibitem [{\citenamefont {See}\ \emph {et~al.}(2012)\citenamefont {See},
  \citenamefont {Pilgrim}, \citenamefont {Scannell}, \citenamefont
  {Montgomery}, \citenamefont {Klochan}, \citenamefont {Burke}, \citenamefont
  {Aagesen}, \citenamefont {Lindelof}, \citenamefont {Farrer}, \citenamefont
  {Ritchie}, \citenamefont {Taylor}, \citenamefont {Hamilton},\ and\
  \citenamefont {Micolich}}]{See12}%
  \BibitemOpen
  \bibfield  {author} {\bibinfo {author} {\bibfnamefont {A.~M.}\ \bibnamefont
  {See}}, \bibinfo {author} {\bibfnamefont {I.}~\bibnamefont {Pilgrim}},
  \bibinfo {author} {\bibfnamefont {B.~C.}\ \bibnamefont {Scannell}}, \bibinfo
  {author} {\bibfnamefont {R.~D.}\ \bibnamefont {Montgomery}}, \bibinfo
  {author} {\bibfnamefont {O.}~\bibnamefont {Klochan}}, \bibinfo {author}
  {\bibfnamefont {A.~M.}\ \bibnamefont {Burke}}, \bibinfo {author}
  {\bibfnamefont {M.}~\bibnamefont {Aagesen}}, \bibinfo {author} {\bibfnamefont
  {P.~E.}\ \bibnamefont {Lindelof}}, \bibinfo {author} {\bibfnamefont
  {I.}~\bibnamefont {Farrer}}, \bibinfo {author} {\bibfnamefont {D.~A.}\
  \bibnamefont {Ritchie}}, \bibinfo {author} {\bibfnamefont {R.~P.}\
  \bibnamefont {Taylor}}, \bibinfo {author} {\bibfnamefont {A.~R.}\
  \bibnamefont {Hamilton}}, \ and\ \bibinfo {author} {\bibfnamefont {A.~P.}\
  \bibnamefont {Micolich}},\ }\href@noop {} {\bibfield  {journal} {\bibinfo
  {journal} {Phys. Rev. Lett.}\ }\textbf {\bibinfo {volume} {108}},\ \bibinfo
  {pages} {196807} (\bibinfo {year} {2012})}\BibitemShut {NoStop}%
\bibitem [{\citenamefont {Mak}\ \emph {et~al.}(2010)\citenamefont {Mak},
  \citenamefont {Gupta}, \citenamefont {Beere}, \citenamefont {Farrer},
  \citenamefont {Sfigakis},\ and\ \citenamefont {Ritchie}}]{Wendy10}%
  \BibitemOpen
  \bibfield  {author} {\bibinfo {author} {\bibfnamefont {W.~Y.}\ \bibnamefont
  {Mak}}, \bibinfo {author} {\bibfnamefont {K.~D.}\ \bibnamefont {Gupta}},
  \bibinfo {author} {\bibfnamefont {H.~E.}\ \bibnamefont {Beere}}, \bibinfo
  {author} {\bibfnamefont {I.}~\bibnamefont {Farrer}}, \bibinfo {author}
  {\bibfnamefont {F.}~\bibnamefont {Sfigakis}}, \ and\ \bibinfo {author}
  {\bibfnamefont {D.~A.}\ \bibnamefont {Ritchie}},\ }\href@noop {} {\bibfield
  {journal} {\bibinfo  {journal} {Appl. Phys. Lett.}\ }\textbf {\bibinfo
  {volume} {97}},\ \bibinfo {pages} {242107} (\bibinfo {year}
  {2010})}\BibitemShut {NoStop}%
\bibitem [{\citenamefont {Sarkozy}\ \emph {et~al.}(2007)\citenamefont
  {Sarkozy}, \citenamefont {Gupta}, \citenamefont {Sfigakis}, \citenamefont
  {Farrer}, \citenamefont {Beere}, \citenamefont {Harrell}, \citenamefont
  {Ritchie},\ and\ \citenamefont {Jones}}]{Sarkozy07}%
  \BibitemOpen
  \bibfield  {author} {\bibinfo {author} {\bibfnamefont {S.}~\bibnamefont
  {Sarkozy}}, \bibinfo {author} {\bibfnamefont {K.~D.}\ \bibnamefont {Gupta}},
  \bibinfo {author} {\bibfnamefont {F.}~\bibnamefont {Sfigakis}}, \bibinfo
  {author} {\bibfnamefont {I.}~\bibnamefont {Farrer}}, \bibinfo {author}
  {\bibfnamefont {H.~E.}\ \bibnamefont {Beere}}, \bibinfo {author}
  {\bibfnamefont {R.}~\bibnamefont {Harrell}}, \bibinfo {author} {\bibfnamefont
  {D.~A.}\ \bibnamefont {Ritchie}}, \ and\ \bibinfo {author} {\bibfnamefont
  {G.~A.~C.}\ \bibnamefont {Jones}},\ }\href@noop {} {\bibfield  {journal}
  {\bibinfo  {journal} {Electrochemical Soc Proc.}\ }\textbf {\bibinfo {volume}
  {11}},\ \bibinfo {pages} {75} (\bibinfo {year} {2007})}\BibitemShut {NoStop}%
\bibitem [{Note2()}]{Note2}%
  \BibitemOpen
  \bibinfo {note} {The timescale for this process is of order of a several
  minutes, with factors being the resistance and capacitance across the AlGaAs
  barrier (thus determining an {$RC$} time constant), and the speed at which
  the overall topgate is ramped (or if held steady). In this scenario, we
  emphasize that no leakage is observed between the 2DEG and the overall
  topgate: no current flows through the insulating polyimide
  layer.}\BibitemShut {Stop}%
\bibitem [{\citenamefont {Gold}(1988)}]{Gold88}%
  \BibitemOpen
  \bibfield  {author} {\bibinfo {author} {\bibfnamefont {A.}~\bibnamefont
  {Gold}},\ }\href@noop {} {\bibfield  {journal} {\bibinfo  {journal} {Phys.
  Rev. B}\ }\textbf {\bibinfo {volume} {38}},\ \bibinfo {pages} {10798}
  (\bibinfo {year} {1988})}\BibitemShut {NoStop}%
\bibitem [{\citenamefont {MacLeod}\ \emph {et~al.}(2009)\citenamefont
  {MacLeod}, \citenamefont {Chan}, \citenamefont {Martin}, \citenamefont
  {Hamilton}, \citenamefont {See}, \citenamefont {Micolich}, \citenamefont
  {Aagesen},\ and\ \citenamefont {Lindelof}}]{MacLeod09}%
  \BibitemOpen
  \bibfield  {author} {\bibinfo {author} {\bibfnamefont {S.~J.}\ \bibnamefont
  {MacLeod}}, \bibinfo {author} {\bibfnamefont {K.}~\bibnamefont {Chan}},
  \bibinfo {author} {\bibfnamefont {T.~P.}\ \bibnamefont {Martin}}, \bibinfo
  {author} {\bibfnamefont {A.~R.}\ \bibnamefont {Hamilton}}, \bibinfo {author}
  {\bibfnamefont {A.}~\bibnamefont {See}}, \bibinfo {author} {\bibfnamefont
  {A.~P.}\ \bibnamefont {Micolich}}, \bibinfo {author} {\bibfnamefont
  {M.}~\bibnamefont {Aagesen}}, \ and\ \bibinfo {author} {\bibfnamefont
  {P.~E.}\ \bibnamefont {Lindelof}},\ }\href@noop {} {\bibfield  {journal}
  {\bibinfo  {journal} {Phys. Rev. B}\ }\textbf {\bibinfo {volume} {80}},\
  \bibinfo {pages} {035310} (\bibinfo {year} {2009})}\BibitemShut {NoStop}%
\bibitem [{\citenamefont {See}\ \emph {et~al.}(2010)\citenamefont {See},
  \citenamefont {Klochan}, \citenamefont {Hamilton}, \citenamefont {Micolich},
  \citenamefont {Aagesen},\ and\ \citenamefont {Lindelof}}]{See10}%
  \BibitemOpen
  \bibfield  {author} {\bibinfo {author} {\bibfnamefont {A.~M.}\ \bibnamefont
  {See}}, \bibinfo {author} {\bibfnamefont {O.}~\bibnamefont {Klochan}},
  \bibinfo {author} {\bibfnamefont {A.~R.}\ \bibnamefont {Hamilton}}, \bibinfo
  {author} {\bibfnamefont {A.~P.}\ \bibnamefont {Micolich}}, \bibinfo {author}
  {\bibfnamefont {M.}~\bibnamefont {Aagesen}}, \ and\ \bibinfo {author}
  {\bibfnamefont {P.~E.}\ \bibnamefont {Lindelof}},\ }\href@noop {} {\bibfield
  {journal} {\bibinfo  {journal} {Appl. Phys. Lett.}\ }\textbf {\bibinfo
  {volume} {96}},\ \bibinfo {pages} {112104} (\bibinfo {year}
  {2010})}\BibitemShut {NoStop}%
\bibitem [{\citenamefont {Klochan}\ \emph {et~al.}(2010)\citenamefont
  {Klochan}, \citenamefont {Chen}, \citenamefont {Micolich}, \citenamefont
  {Hamilton}, \citenamefont {Muraki},\ and\ \citenamefont
  {Hirayama}}]{Klochan10}%
  \BibitemOpen
  \bibfield  {author} {\bibinfo {author} {\bibfnamefont {O.}~\bibnamefont
  {Klochan}}, \bibinfo {author} {\bibfnamefont {J.~C.~H.}\ \bibnamefont
  {Chen}}, \bibinfo {author} {\bibfnamefont {A.~P.}\ \bibnamefont {Micolich}},
  \bibinfo {author} {\bibfnamefont {A.~R.}\ \bibnamefont {Hamilton}}, \bibinfo
  {author} {\bibfnamefont {K.}~\bibnamefont {Muraki}}, \ and\ \bibinfo {author}
  {\bibfnamefont {Y.}~\bibnamefont {Hirayama}},\ }\href@noop {} {\bibfield
  {journal} {\bibinfo  {journal} {Appl. Phys. Lett.}\ }\textbf {\bibinfo
  {volume} {96}},\ \bibinfo {pages} {092103} (\bibinfo {year}
  {2010})}\BibitemShut {NoStop}%
\bibitem [{\citenamefont {Foxman}\ \emph {et~al.}(1994)\citenamefont
    {Foxman},
  \citenamefont {Meirav}, \citenamefont {McEuen}, \citenamefont {Kastner},
  \citenamefont {Klein}, \citenamefont {Belk}, \citenamefont {Abusch},\ and\
  \citenamefont {Wind}}]{Foxman94}%
  \BibitemOpen
  \bibfield  {author} {\bibinfo {author} {\bibfnamefont {E.~B.}\ \bibnamefont
  {Foxman}}, \bibinfo {author} {\bibfnamefont {U.}~\bibnamefont {Meirav}},
  \bibinfo {author} {\bibfnamefont {P.~L.}\ \bibnamefont {McEuen}}, \bibinfo
  {author} {\bibfnamefont {M.~A.}\ \bibnamefont {Kastner}}, \bibinfo {author}
  {\bibfnamefont {O.}~\bibnamefont {Klein}}, \bibinfo {author} {\bibfnamefont
  {P.~A.}\ \bibnamefont {Belk}}, \bibinfo {author} {\bibfnamefont {D.~M.}\
  \bibnamefont {Abusch}}, \ and\ \bibinfo {author} {\bibfnamefont {S.~J.}\
  \bibnamefont {Wind}},\ }\href@noop {} {\bibfield  {journal} {\bibinfo
  {journal} {Phys. Rev. B}\ }\textbf {\bibinfo {volume} {50}},\ \bibinfo
  {pages} {14193} (\bibinfo {year} {1994})}\BibitemShut {NoStop}%
\bibitem [{\citenamefont {Foxman}\ \emph {et~al.}(1993)\citenamefont
    {Foxman},
  \citenamefont {McEuen}, \citenamefont {Meirav}, \citenamefont {Wingreen},
  \citenamefont {Meir}, \citenamefont {Belk}, \citenamefont {Belk},
  \citenamefont {Kastner},\ and\ \citenamefont {Wind}}]{Foxman93}%
  \BibitemOpen
  \bibfield  {author} {\bibinfo {author} {\bibfnamefont {E.~B.}\ \bibnamefont
  {Foxman}}, \bibinfo {author} {\bibfnamefont {P.~L.}\ \bibnamefont {McEuen}},
  \bibinfo {author} {\bibfnamefont {U.}~\bibnamefont {Meirav}}, \bibinfo
  {author} {\bibfnamefont {N.~S.}\ \bibnamefont {Wingreen}}, \bibinfo {author}
  {\bibfnamefont {Y.}~\bibnamefont {Meir}}, \bibinfo {author} {\bibfnamefont
  {P.~A.}\ \bibnamefont {Belk}}, \bibinfo {author} {\bibfnamefont {N.~R.}\
  \bibnamefont {Belk}}, \bibinfo {author} {\bibfnamefont {M.~A.}\ \bibnamefont
  {Kastner}}, \ and\ \bibinfo {author} {\bibfnamefont {S.~J.}\ \bibnamefont
  {Wind}},\ }\href@noop {} {\bibfield  {journal} {\bibinfo  {journal} {Phys.
  Rev. B}\ }\textbf {\bibinfo {volume} {47}},\ \bibinfo {pages} {10020}
  (\bibinfo {year} {1993})}\BibitemShut {NoStop}%
\bibitem [{\citenamefont {Beenakker}(1991)}]{Beenakker91}%
  \BibitemOpen
  \bibfield  {author} {\bibinfo {author} {\bibfnamefont {C.~W.~J.}\
  \bibnamefont {Beenakker}},\ }\href@noop {} {\bibfield  {journal} {\bibinfo
  {journal} {Phys. Rev. B}\ }\textbf {\bibinfo {volume} {44}},\ \bibinfo
  {pages} {1646} (\bibinfo {year} {1991})}\BibitemShut {NoStop}%
\end{thebibliography}

%

\end{document}